%% file: paper.tex
\documentclass[9pt,singlecolumn]{article}

\usepackage{amsmath,amsgen,amstext,amssymb}
\usepackage{amsfonts}
\usepackage{bbm}
\usepackage{latexsym}
\usepackage[english]{babel}
\usepackage{cite}
\usepackage{graphicx}
\usepackage{tikz}
\usetikzlibrary{calc}
\usepackage{color}
\usepackage{booktabs}
\usepackage[caption=false,font=footnotesize]{subfig}
\usepackage{appendix}
\usepackage[numbers]{natbib}
\usepackage{balance}

\newcommand{\comp}{\hbox{ o }}
\def\done{\hspace*{\fill} \rule{1.8mm}{2.5mm}}

  \renewenvironment{thebibliography}[1]{%
    \begin{oldthebibliography}{#1}%
      \setlength{\parskip}{0ex}%
      \setlength{\itemsep}{0ex}%
  }%
  {%
    \end{oldthebibliography}%
  }

\newcommand{\cA}{\mathcal{A}}

\newcommand{\cB}{\mathcal{B}}

\newcommand{\vbeta}{\vec{\beta}}

\newcommand{\baq}{\begin{eqnarray}}
\newcommand{\eaq}{\end{eqnarray}}

\newcommand{\baqm}{\begin{eqnarray*}}
\newcommand{\eaqm}{\end{eqnarray*}}

\newcommand{\beq}{\begin{equation}}
\newcommand{\eeq}{\end{equation}}

\def\boldeta {\mbox{\boldmath $\eta$}}  

\newcommand{\bone}{\boldsymbol{1}}

\newenvironment{pf}{{\bf Proof. }}{\hfill $\square$\medskip}
\newtheorem{theo}{Theorem}[section]
\newtheorem{ass}{Assumption}[section]
\newtheorem{defi}{Definition}[section]
\newtheorem{prop}{Proposition}[section]

\newtheorem{lem}{Lemma}[section]

\include{definitions}

\newcommand{\domath}[2]{%
\pgfmathparse{#1}%
\pgfmathsetmacro#2{\pgfmathresult}}

\begin{document}
\title{Characterizing Continuous Time Random Walks \\ on Time Varying Graphs
\thanks{\scriptsize
This work was supported by the NSF grant CNS-1065133 and ARL Cooperative Agreement W911NF-09-2-0053. 
The views and conclusions contained
in this document are those of the authors and should not be interpreted as
representing the of official policies, either expressed or implied of the NSF,
ARL, or the U.S. Government. The U.S. Government is authorized
to reproduce and distribute reprints for Government purposes notwithstanding
any copyright notation hereon.  The work of E. de Souza e Silva and D. Figueiredo was
partially supported by grants from CNPq and Faperj. 
}
}

\date{Technical Report UM-CS-2012-011v2}

\author{
{	D.\ Figueiredo$^1$, P.\ Nain$^2$, B. Ribeiro$^3$, E.\ de Souza e Silva$^1$ and D.\ Towsley$^3$ }\\
~ \\
\centering \normalsize
       \begin{tabular}[t]{ccc}
       $^1$ Systems Engineering Department &  $^2$ INRIA & $^3$ Department of Computer Science   \\
	COPPE                             & B.P. 93 & University of Massachusetts Amherst \\
	Federal University of Rio de Janeiro  &06902 Sophia Antipolis  & 140 Governors Drive   \\
	Rio de Janeiro, Brazil  & France & Amherst, MA 01003 \\
	 \{daniel,edmundo\}@land.ufrj.br & philippe.nain@inria.fr & \{ribeiro,towsley\}@cs.umass.edu
        \end{tabular}
}

\renewcommand\footnotemark{}


\maketitle

\begin{abstract}
In this paper we study the behavior of a continuous time random walk
(CTRW) on a stationary and ergodic time varying dynamic graph.  We establish conditions under
which the CTRW is a stationary and ergodic process.  In general, the stationary
distribution of the walker depends on the walker rate and is difficult
to characterize. However, we characterize the stationary distribution in 
the following cases: i) the walker rate is significantly larger or smaller 
than the rate in which the graph changes (time-scale separation), ii) the 
walker rate is proportional to the degree of the node that it resides on (coupled 
dynamics), and iii) the degrees of node belonging to the same connected 
component are identical (structural constraints). We provide
examples that illustrate our theoretical findings.
\end{abstract}



\input{introduction}

\input{model}


\section{Characterizing RWs: Stationary Behavior}
\label{sec:SS}
In this section we focus on the stationary behavior of a RW on a dynamic graph, in particular 
the steady state fraction of time the walker spends in each node of the network: 
$\bpi = (\pi_1, \ldots, \pi_n)$. 

The steady state distribution $\bpi$ is trivial if the graph is static and T-connected, i.e., $A(t)=A^\prime$, $\forall t \geq 0$, where $A^\prime$ is a symmetric $(0,1)$-adjacency matrix of a connected graph. The stationary distribution is unique and given by \[ \pi_v = \frac{d_v/\gamma_v}{\sum_{w} d_w/\gamma_w}, \quad v\in V, \]
where $d_v$ denotes the degree of node $v$ and $\gamma_v$ is the walker rate at node $v$.
Characterization on a dynamic graph is much more challenging.
How can we characterize $\bpi$ on dynamic graphs? When does $\bpi$ converge for dynamic graphs? When can we give an expression for $\bpi$?   We focus on three cases where we are able to characterize this behavior.

\input{stationary}

\subsection{Stationary Behavior under Time-scale Separation} 
Consider a scenario where the walker is either much faster or much slower relative to the rate that the graph changes configurations.  In this case, we have a time-scale separation between the two processes that allows us to characterize $\bpi$.

\input{fast}


\input{slow_alt-new}

\input{stable_walker}

\input{casestudies}

\input{related}

\input{conclusion}

%
\balance
{
\bibliographystyle{plain}

\input{paper.bbl}
}

\section*{Appendices}
\input{appendix}

\end{document}

%% file: definitions.tex
\newcommand{\bP}{ {\bf P} }

\newcommand{\bpi}{ {\mbox{\boldmath $\pi$}} }

\newcommand{\bpsi}{ {\mbox{\boldmath $\psi$}} }

\newcommand{\bgamma}{ {\mbox{\boldmath $\gamma$}} }

\newcommand{\bsigma}{ {\mbox{\boldmath $\sigma$}} }

\newcommand{\bc}{ {\bf c} }

\newcommand{\be}{ {\bf e} }

\newcommand{\vgamma}{ {\mbox{$\vec{\bgamma}$}} }

%% file: introduction.tex
\section{Introduction}

During the last decade, there has been a wide interest in characterizing and 
modeling the structure of various networks, from neural networks, to the web, to Facebook friends. Real 
networks are inherently dynamic in the sense that both nodes and edges come and go 
over some time-scale. However, most efforts consider the network as either a 
single static graph or as a pre-defined sequence of graph configurations. 

Random walks are an important building blocks for characterizing networks. Their simple 
behavior on static networks has been explored to devise algorithms for various purposes, from ranking to 
searching (details in Section~\ref{sec:related}). However, very little is known about the long-term behavior 
of random walks on dynamic networks.

In this paper, we study continuous time random walks (CTRWs) on stationary and ergodic dynamic graphs. 
We make the following contributions towards this goal:
\begin{itemize}
\item
We consider stationary and ergodic dynamic graphs where nodes are always present in the network but 
edges are allowed to come and go over time, including the cases where the network consists of several
connected components. We introduce the notion of T-connectivity and show that if the dynamic graph is 
stationary, ergodic and T-connected then the CTRW is also stationary and ergodic. In the full generality 
of our framework, the stationary distribution of the walker depends on the walker rate and is difficult
to characterize. However,

\item 
we characterize the stationary distribution of the random walk for several cases: {\it (i)} {\bf Time-scale separation}: 
the walker rate is significantly larger or smaller than the rate in which the graph changes; {\it (ii)} {\bf Coupled 
dynamics}: the walker rate is proportional to the degree of the node that it resides on; {\it (iii)} {\bf Structural 
constraints}: the degrees of nodes within any connected components are identical (but can vary among 
different components).

\item
We evaluate numerically several examples to support our theoretical results and illustrate their 
applicability. We also present a simple illustrative DTN application. 

\end{itemize}

The remainder of this paper is organized as follows. Section~\ref{sec:model} presents the proposed modeling 
framework together with some definitions and properties. Section~\ref{sec:SS} presents the stationary distribution 
of CTRW when the walker rate (for being too fast or too slow in respect to the speed the graph changes) allows a 
time scale decomposition of the combined walker and graph processes;
we also present conditions under which the CTRW stationary distribution is invariant under time scale changes.
Section~\ref{sec:casestudies} presents the numerical examples and 
applications. In Section~\ref{sec:related} we discuss the related work. Finally, Section~\ref{sec:conclusion} 
concludes the paper.

%% file: model.tex
\section{Model formulation}
\label{sec:model}

In this section we define important concepts that will be used throughout this paper.
We define a dynamic graph as a simple marked point process.
A continuous time random walk (CTRW) over a dynamic graph is also a marked point process.
We define the concept of T-connectivity to express the ability of nodes to be connected in time.
We also define conditions for stationarity of the graph process and the CTRW process.
We start by defining the graph dynamics.

\begin{defi} [Dynamic Graph] \label{def:G}
The time evolution of the graph under consideration is given by the (possibly simultaneous) addition and deletion of edges.
Let $V$ denote a finite set of $n$ nodes and let $\cA$ denote a finite set of $m$ adjacency matrices $\cA = \{A_k\}_{k=1}^m$, where $A_k$ is an $n \times n$  unweighted symmetric adjacency matrix.
A dynamic graph is a simple Random Marked Point Process (RMPP) $\Psi = \{(X_i,S_i)\}_{i \in \mathbb{Z}}$,  where $\mathbb{Z}$ is the set of all integers, $X_i \in \cA$ denotes the $i$-th graph configuration and $S_i$ is the time that the network spends in that configuration. 
\end{defi}

Because $\Psi$ is simple, $P(S_i =0) = 0$ for all $i$. 
Moreover, we assume $0 < E[S_i]< \infty$, $i \in \mathbb{Z}$. 
We use $G_k$ to denote the graph configuration that has adjacency matrix $A_k$, $k=1,\dots,m$.
Throughout this work we use $A_k$ and $G_k$ interchangeably.
To simplify our analysis we focus on unweighted adjacency matrices.
However, edge weights can be easily accounted for in the walker rates and thus our results are also applicable to weighted dynamic graphs.
We define the process $\{A(t)\}_{t \in \mathbb{R}}$, as 
\begin{equation} \label{eq:At}
A(t) = \sum_{i \in \mathbb{Z}} X_i \bone_{\{T_i \leq t < T_{i+1}\}}  \,,
\end{equation}
where by convention
$\cdots < T_{-1} < T_0 \leq 0 < T_1  <\cdots$ are the successive times  at which the graph switches 
to another configuration with $S_i:=T_{i+1}-T_i$, $i \in \mathbb{Z}$,  
i.e., $A(t)$ denotes the adjacency matrix of the graph at time $t$; $A(t)$ is right-continuous. 

 \begin{ass}
\label{ass:stat-erg}
The graph process $\Psi$ is stationary and ergodic.
\end{ass}
Throughout this paper we assume that Assumption~\ref{ass:stat-erg}  holds.
The following proposition shows that the ergodicity and stationarity of $\Psi = \{X_i,S_i\}_{i \in \mathbb{Z}}$ implies $\{A(t)\}_{t \in \mathbb{R}}$ is also stationary and ergodic.
\begin{prop}\label{prop:AtSE}
Under Assumption \ref{ass:stat-erg}, $\{A(t)\}_{t \in \mathbb{R}}$ is stationary and ergodic.
\end{prop}
A proof of Proposition~\ref{prop:AtSE} is provided in Appendix~\ref{stationarity}. In what follows we provide a result that will be useful in our proofs:
\begin{theo} \label{teo:birkhoff}
Under Assumption \ref{ass:stat-erg},
\begin{equation}\label{eq:birkhoff}
\lim_{t\to\infty}\frac{1}{t}\int_0^t \bone_{\{A(x)=A_k\}} dx =\sigma_k \, , \quad \hbox{P-a.s.,}
\end{equation}
where $\sigma_k > 0$.
\end{theo}
\begin{pf}
The proof of Theorem~\ref{teo:birkhoff} follows directly from Proposition~\ref{prop:AtSE} and Birkhoff's ergodic theorem (Shiryaev~\cite[pg. 409, Theorem 1]{Shiryaev}). 
\end{pf}

We refer to $\sigma = (\sigma_1,\ldots , \sigma_m)$ as the stationary distribution of $\{A(t)\}_{t \geq 0}$.
We note in passing that a direct consequence of Theorem~\ref{teo:birkhoff} is that $\lim_{t\to\infty}\frac{1}{t}\int_0^t \bone_{\{A(x)=A_k\}} dx =  E[ \bone_{\{X_i=A_k\}} S_i] / E[S_i] = \sigma_k$, for all $i,k$.
 
Another important property of a stationary and ergodic $\Psi$ that we use throughout this work is the uniform convergence of the time average regardless of the initial conditions. 
\begin{prop} [Uniform mixing] \label{prop:uniform} 
Under Assumption \ref{ass:stat-erg},
\begin{equation}
\lim_{t\to\infty} \frac{1}{t}\int_0^t  P(A(x) = A_k | \cB)dx = \sigma_k \, , \quad k=1,\ldots,m\, ,
\label{assumption-mixing-uniform}
\end{equation}
for any Borel set $\cB$ such that $P(\cB)>0$. Furthermore, the convergence is uniform with respect to any $\cB$.
\end{prop}
\begin{pf}
A short proof is presented. An extended proof can be found at Appendix~\ref{uniform}.
Stationarity and ergodicity imply~\eqref{eq:birkhoff}.
Integrating both sides of~\eqref{eq:birkhoff} w.r.t.\  $dP(\omega)$ for $\omega \in \cB$ and using the bounded convergence theorem,
Fubini's theorem and Egorov's theorem~\cite[pg.\ 43, Theorem 2]{Kolmogorov} gives~\eqref{assumption-mixing-uniform} 
uniformly in $\cB$ for any Borel set $\cB$ such that $P(\cB)>0$.
\end{pf}

It is possible for one or more graph configurations to consist of two or more disconnected components, in which case we have to be concerned about whether a walker can move from any node to any other node over time.  In what follows we introduce the concept of connectivity over time between two nodes (denoted as T-connectivity). 
\begin{defi} [T-connectivity] \label{def:Tc}
Two nodes $u$ and $v$ are said to be T-connected in $\Psi$, if they are connected in the graph induced by adjacency matrix $A =  \vee_k A_k$, where $\vee$ is the binary OR operator.
\end{defi}
An alternative equivalent definition of T-connectivity uses  $A =  \sum_{k=1}^m A_k$ in place of $A =  \vee_k A_k$.
T-connectivity can also be stated as a graph property.
\begin{defi} [T-connected graph]
A dynamic graph is said to be T-connected if all pairs of nodes are T-connected.
\end{defi}
%
We now define a Continuous Time Random Walk (CTRW) over the dynamic graph, starting at time $t = 0$.
\begin{defi} [CTRW]
A continuous time random walk (CTRW) on a dynamic graph $\{A(t)\}_{t \in \mathbb{R}}$ associated with RMPP $\Psi$, is a process $\{(A(t),U(t))\}_{t \geq 0}$, where $U(t)\in V$ is the position (node) of the walker at time $t$. The times between CTRW steps are independent and exponentially distributed.
The rate at which the walker makes a step at node $U(t)=v$ when $A(t)=A_k$ is $\gamma_{k,v}$. 
At the time the walker leaves $v$, it chooses one of its currently connected neighbors in $A_k$ (if any) uniformly at random. When $v$ has no neighbors in $A_k$ the walker stays at $v$ until the next step event.
\end{defi}
Let $\Gamma=\{\gamma_{k,v}\}$ denote the set of walker rates associated with $\{A(t)\}_{t \geq 0}$. We will find it useful to express $\gamma_{k,v}$ as $\gamma_{k,v} = \beta_{k,v} \gamma$, $k=1,\dots,m$ and $v \in V$.
Walker rates of interest to us include $\beta_{k,v} = 1$ (denoted {\em CTRW with constant walker rate}) and $\beta_{k,v} =  d_{k,v}$, where $d_{k,v}$ is the degree of $v$ given adjacency matrix $A_k$ (denoted {\em CTRW with degree dependent walker rate}). 

The above framework is general enough to describe several more 
particular dynamic graph models, such as renewal processes and Markovian processes.
In a Markovian process $S_i$ is exponentially distributed and
$P[X_i=x_i | X_{i-1}=x_{i-1},X_{i-2}=x_{i-2},\dots ] = P[X_i=x_i | X_{i-1}=x_{i-1}]$, $x_i\in V$, $i=0,1,\dots$.

\begin{center}
\begin{tabular}{@{}ll@{}} 
\multicolumn{2}{c}{\bf Notation Summary} \\
\toprule
$\Psi=\{(X_i,S_i)\}_{i \in \mathbb{Z}}$  & dynamic graph process (in events)\\
$\{A(t)\}_{t \in\mathbb{R}}$ & dynamic graph process (in time) \\
$\{(A(t),U(t))\}_{t \geq 0}$ & CTRW process \\
$\sigma = (\sigma_1,\ldots , \sigma_m)$ & stationary distribution of $\{A(t)\}_{t \geq 0}$ \\
$\gamma_{k,v} = \beta_{k,v} \gamma$ & {\em walker rate} of the CTRW at node\\ & $v$ at configuration $A_k$\\
\bottomrule
\end{tabular}
\end{center}

%% file: stationary.tex
We begin our study of the stationary behavior of the process $\{U(t)\}$ with the following.
\begin{theo} \label{CTRW-stionary}
If the RMPP $\Psi$ is
$\mbox{T-connected}$, stationary, and ergodic, then the process $\{(A(t),U(t))\}_{t \geq 0}$ is stationary, and the stationary distribution is unique.
\end{theo}
\begin{pf}
We create a new marked point process, $\{(X'_i,S'_i)\}_0^\infty$  that is a superposition of the graph process  $\{(X_j,S_j)\}_0^\infty$ and a Poisson process having rate $\gamma_{\max} = \max_{r \in \Gamma} r$ (recall that $\Gamma$ is the set of walker rates). 
To simplify our proof we shall assume $P[T_0 = 0] = 1$, where $T_0$ is as defined in~\eqref{eq:At}, although the proof is valid for any $T_0 \leq 0$.
We associate the mark ``0'' with each point of the Poisson process,  Hence $X'_i \in \{0\}  \cup \{A_1, \ldots , A_m\}$.   As both the graph and Poisson processes are event stationary, the new merged process is also event stationary \cite[Section 1.3.5]{Franken}. Let $t'_0 < t'_1 < \cdots < t'_i\le \cdots$ denote the times associated with this new process, $\{(X'_i,S'_i)\}_0^\infty$.  Consider the process $\{(A'_i,U_i)\}_{0}^\infty$ where $U_i$ denotes the walker position at  time $t'_i$ and $A'_i$ denotes the adjacency matrix during the period $[t'_i,t'_{i+1})$.   Note that $(A'_i,U_i) \in \cA \times V$ takes values from a finite set.  $\{(A'_i,U_i)\}$ is described by a stochastic recursion of the form $(A'_i,U_i) = \phi (U_{i-1}, X'_i, R_i)$ where  
\baqm
A'_i & = &  \phi_a(U'_{i-1}, X'_i, R_i) = \bone\{X'_i = 0\}A'_{i-1} + \bone\{X'_i \ne 0\} X'_i, \\
U_i & = & \phi_b (U'_{i-1}, X'_i, R_i),
\eaqm
for all $i = 0, \ldots$. Here  $X'_i, S'_i$ are as previously defined and $\{R_i\}$ is an iid  sequence of uniformly distributed rvs in $[0,1]$ independent of $\{(X'_i,S'_i)\}$. These auxiliary rvs are used to choose the neighbor to which the walker goes or to remain stationed at its current node.  Note that $\{(X'_i,S'_i,R_i)\}$ is stationary.  $\phi_b$ is defined so that when $X'_i \ne 0$, the walker does not move ($U_i = U_{i-1}$) but the graph changes to configuration $X'_i$.  If $X'_i = 0$, the walker moves from $U_i$ with probability $\gamma_{X'_i,U_i}/\gamma_{\max}$, moving to one of its neighbors (in configuration $A'_{i-1}$) chosen uniformly at random (using $R_i$).

Theorem 1 in \cite{moyal} states that if there exists a random subset, $B \subseteq\cA\times V$ such that a sample path monotonicity condition ((5) in \cite{moyal}) holds and the existence of a finite non-empty sample path absorbing set ((9) in \cite{moyal}) exists, then it is possible to construct process $\{X'_i,S'_i,A'_i,U_i\}$ that is event stationary as is $\{U_i\}$.  In our case, because our state space is finite, these conditions trivially hold by taking $B = \cA\times V$.  
Since $\{X'_i,S'_i,A'_i, U_i\}$ is event stationary, $\{A'(t),U(t)\}$
is also stationary. It follows from our construction that $\{A(t)\} = \{A'(t)\}$; hence $\{A(t),U(t)\}$ is also stationary. 

We now address the question of uniqueness through a coupling argument.   Consider two random walks $\{\{U_1(t)\}_{t \geq 0}$ and $\{U_2(t)\}\}_{t \geq 0}$ that differ in their starting locations at time $t=0$, $U_1(0) = u_1$ and $U_2(0) = u_2$.  We are interested in establishing that the time, $T$ at which they meet, is finite a.s..  After time $T$ the processes couple, i.e., for $t > T$, $U_1(t) = U_2(t)$. This is possible because the times between steps are exponentially distributed random variables.
Thus, when $T$ is finite, the above coupling argument implies that $\{U_1(t)\}$ and $\{U_2(t)\}$, which we have shown to be time asymptotic stationary, have the same stationary distribution.

It is left to show is that $T$ is finite.  We sketch the argument here and relegate the details to Appendix~\ref{largeT}.  The basic idea is to identify intervals of time of length $T^\prime <\infty$ starting at times $iT^\prime \geq 0$, $i=0,1,\ldots$, and based on the ergodicity and time stationarity of the graph process to establish a lower bound, $p_0$, on the probability of two walkers coupling during interval $[iT^\prime,(i+1)T^\prime]$. The probability that the walkers do not couple within the interval $[0,jT_0)$ is upper bounded by $(1-p_0)^j$.  Thus the walkers couple in finite time a.s..
%
\end{pf}

Note that if the graph process is not T-connected, then it is possible for the system to exhibit multiple stationary regimes that depend on the initial position of the walker.
Next we characterize $\bpi$ when there is a time-scale separation of the walker and graph dynamics.

%% file: fast.tex
\subsubsection{The Fast Walker}\label{sec:fast}
Let us first assume that the walker rate is much larger than the rate at which the graph changes.     
For a sufficiently large $\gamma$, the 
steady state probabilities of the random walk $\bpi$ is a linear combination of
the corresponding probabilities of the adjacency matrices $A_1,\ldots ,A_m$.
Theorem~\ref{th:fast-w} formalizes this argument for the case that every adjacency matrix in $\cA$ is connected.  
We will describe, under certain conditions, how to relax this assumption later.

In preparation, let $\gamma_{k,v} = \beta_{k,v} \gamma$ and let  $\bpi^{(k)} (\gamma) = ( \pi^{(k)}_1(\gamma),$ $ \ldots, \pi^{(k)}_n(\gamma) )$ denote 
the steady state distribution of a random walk on the undirected graph with adjacency matrix $A_k$ as a function of $\gamma >0$. 
It is given by 
\beq \label{eq:stationary_u}
\pi^{(k)}_v(\gamma) \equiv \pi^{(k)} = \frac{d_{k,v}/\beta_{k,v}}{\sum_{j\in V} d_{k,j}/\beta_{k,j}}, \quad v\in V; k=1,\ldots , m. 
\eeq
independent of $\gamma$.
Let
\[
\bpi^{(k)}(\gamma, t,w) = (\pi^{(k)}_1(\gamma, t,w), \ldots, \pi^{(k)}_n(\gamma, t,w))
\]
denote the distribution of the CTRW on $A_k$ at time $t \ge 0$ starting from node $w$. 
Assume $A_k$ is irreducible, $k=1,\ldots,m$, i.e., the graph with adjacency matrix $A_k$ is connected.
Because the random walk with adjacency matrix $A_k$ is described by a time-reversible Markov chain, 
$\bpi^{(k)}(\gamma,t,w)$ can be expressed as 
\begin{equation} 
\bpi^{(k)}(\gamma,t,w) = \bpi^{(k)} + \sum_{j=2}^n \bc^{(k)}_{j,w} e^{\lambda_{kj}\gamma t},\quad w\in V, \; t>0 \, ,
\label{reversible}
\end{equation}
where $0 = \lambda_{k1} > \lambda_{k2}\ge \cdots \ge \lambda_{km}$ 
are the eigenvalues associated with $Q_k(\gamma)/\gamma$ where $Q_k(\gamma)$ is the infinitesimal generator associated with the random walk with parameter $\gamma$ on the graph with adjacency matrix $A_k$,
and $\{\bc^{(k)}_{j,w}\}$ are vectors  related to the $j$-th eigenvector of the random walk and the initial condition that
the walker begins at node $w$.
\begin{theo}
\label{th:fast-w}
If the graph process $\Psi$ is T-connected, stationary, ergodic, and the configurations are always connected, then in the limit as $\gamma \rightarrow \infty$, the stationary 
distribution $\bpi$ of the random walk is given by 
\begin{equation}
\label{eq:theor-fast-gamma}
\bpi = \sum_{k=1}^m \sigma_k\bpi^{(k)}.
\end{equation}
\end{theo}
\begin{pf}
We show that the walker steady state distribution $\bpi (\gamma) \rightarrow \bpi$ as $\gamma \rightarrow \infty$ where $\bpi$ is given in (\ref{eq:theor-fast-gamma}).

We focus on the $i$-th graph configuration, $X_i$, $i \geq 0$. 
To simplify our proof we shall assume $P[T_0 = 0] =1$, where $T_0$ is as defined in~\eqref{eq:At}.
Let $F^{(k)}_{i}(x) = P(S_i \le x | X_i = A_k)$ and define $\boldeta^{(k)}_{i}(\gamma)$ to be the stationary distribution of
the CTRW while the graph is in state $X_i = A_k$. 
Let $P_{i,w}(\gamma)$ denote the initial walker distribution when
the process first enters graph configuration $X_i $. 
$\boldeta^{(k)}_{i}(\gamma)$ is defined as
\baqm
\boldeta^{(k)}_{i}(\gamma)
 & = & \sum_{w\in V} P_{i,w}(\gamma) \int_{0}^\infty \frac{1}{t} \int_{0}^t \bpi^{(k)}(\gamma,x,w) dx\; dF^{(k)}_{i} \\
 & = &  \bpi^{(k)}  + \sum_{w\in V} P_{i,w}(\gamma) \int_{0}^\infty \frac{1}{t} \int_{0}^t\sum_{j=2}^n \bc^{(k)}_{w} e^{\lambda_{kj}\gamma x} dx\;
dF^{(k)}_{i}  \, ,
\eaqm
where the second equality follows from (\ref{reversible}).  We focus on the second term, henceforth denoted as $C_\gamma$, which we show goes to zero as $\gamma \rightarrow \infty$. We focus first on the singularity of the $1/t$ term due to the first integral starting from zero,
\baqm
|C_\gamma| < F^{(k)}_{i}(\gamma^{-1/4})\be \mbox{} + \sum_{w\in V} P_{i,w}(\gamma) \int_{\gamma^{-1/4}}^\infty \frac{1}{t} \int_0^t \sum_{j=2}^n |\bc^{(k)}_{w}| e^{\lambda_{kj}\gamma x} dx\; dF^{(k)}_{i} \, ,
\eaqm
where $|\bc|$ is the vector whose components are the absolute values of the components of $\bc$ and $\be$ is a vector of all ones. 
Evaluating the second integral and recognizing that $1/t \le \gamma^{1/4}$ for $t\ge \gamma^{-1/4}$ yields
\baqm
|C_\gamma|  &< &  F^{(k)}_{i}(\gamma^{-1/4})\be
 \mbox{} +{\gamma^{1/4}} \sum_{w\in V} P_{i,w}(\gamma) \sum_{j=2}^n \int_{\gamma^{-1/4}}^\infty  \frac{|\bc^{(k)}_{w}|}{\lambda_{kj}\gamma} (e^{\lambda_{kj}\gamma t}-1)dF^{(k)}_{i} \\
 & \le & F^{(k)}_{i}(\gamma^{-1/4})\be  \mbox{} + \frac{1}{\gamma^{3/4}} \sum_{w\in V} P_{i,w}(\gamma) \sum_{j=2}^n \int_{\gamma^{-1/4}}^\infty  \frac{|\bc^{(k)}_{w}|}{(-\lambda_{kj})} dF^{(k)}_{i} \\
 & \le &  F^{(k)}_{i}(\gamma^{-1/4})\be \nonumber  \mbox{} + \frac{1}{\gamma^{3/4}} \sum_{w\in V} P_{i,w}(\gamma) \sum_{j=2}^n  -\frac{|\bc^{(k)}_{w}|}{\lambda_{kj}} (1- F^{(k)}_{i}(\gamma^{-1/4})) \, .
\eaqm
 Both terms go to zero as $\gamma\rightarrow \infty$.  Consequently $C_\gamma \rightarrow 0$ and
\[ \lim_{\gamma\rightarrow \infty} \boldeta^{(k)}_i(\gamma) = \bpi^{(k)}, \quad \forall k \]

This holds for all $i$; therefore it holds when the graph is in steady state and removal of the conditioning on the graph configuration yields  (\ref{eq:theor-fast-gamma}).
\end{pf}

{
We now focus on the case where one or more of the graph configurations consists of disconnected components. We relabel the nodes in each graph configuration in order to easily identify the disconnected components.   For each of the original $m$ adjacency matrices, $A_k$,  we rearrange the $n$ nodes into subsets of connected components.  In other words, consider graph configuration $G_k$ associated with adjacency matrix $A_k$.  Partition the set of nodes in $G_k$ into $o_k$ sets, each containing only connected nodes.  The $o_k$ sets correspond to $o_k$ adjacency matrices, $\{A_{k,1}, \ldots, A_{k,o_k}\}$ and graph configurations $\{G_{k,1}, \ldots, G_{k,o_k} \}$.  Let $V_{k,l}$ denote the set of nodes in configuration $G_{k,l}$.  

Let $\bpsi (\gamma) = (\psi_{1,1}(\gamma),\ldots , \psi_{m,o_m}(\gamma))$ denote the vector of stationary probabilities that the walker is in the different components of all of the configurations when the rate parameter is $\gamma$.  Because the CTRW process is ergodic, this vector exists and is given by 
\[ 
\psi_{k,l}(\gamma) = \lim_{t\rightarrow \infty} \frac{1}{t} \int_0^t \sum_{v\in V_{k,l}} \bone_{\{A(s) = A_k,U(s) = v\}} ds, \quad \forall k,l \, .
\]
Define  
\begin{equation}\label{eq:bpsi}
\bpsi = \lim_{\gamma \rightarrow \infty} \bpsi (\gamma) \, .
\end{equation}
We will describe conditions under which $\bpsi$ can be computed later.
In what follows we show that if $\bpsi$ exists then we obtain the stationary distribution $\bpi$ of the random walk in the limit as $\gamma \to \infty$.


Let $\bpi^{(k,l)}$ be the steady state distribution of a random walk on the undirected graph
with adjacency matrix $A_{k,l}$.
Similar to equation (\ref{eq:stationary_u})\footnote{If the denominator is zero (i.e., there are isolated nodes) we simplify our notation assuming the ratio $0/0 = 1$.},
\[
\bpi^{(k,l)}_v = \frac {d_{k,v}/\beta_{k,v} } {\sum_{j \in V_{k,l} } d_{k,j}/\beta_{k,j} }, \quad 
v \in V_{k,l} ,
\]
and $k=1,\ldots, m, l=1,\ldots, o_k$.

We let $\bpi^{(k)}$ be the concatenation of vectors $\bpi^{(k,l)}$, that is,
$\bpi^{(k)} = ( \bpi^{(k,1)} \| \ldots \| \bpi^{(k,o_k)} )$
and 
\[
\hat{\bpi}^{(k)} =  ( \psi_{k,1} \bpi^{(k,1)} \| \ldots \| \psi_{k,o_k} \bpi^{(k,o_k)} ) .
\]
Note that vectors $\hat{\bpi}^{(k)}$ for all $1 \leq k \leq m$ have the same cardinality.

We have the following result.
 \begin{theo}
\label{th:fast-w-disc}
If the graph process $\Psi$ is T-connected, stationary, and ergodic, and $\bpsi$ exists, then in the limit as $\gamma \rightarrow \infty$,
the stationary distribution $\bpi$ of 
the random walk when graph configurations may be disconnected is given by
\begin{equation}  
\label{eq:theor-fast-gamma-disc}
\bpi = \sum_{k=1}^m \hat{\bpi}^{(k)} .
\end{equation}
\end{theo}
\begin{pf} The proof is similar to that for the case where all graphs are connected.  
\end{pf}

In general $\bpsi$ is difficult to compute.
The difficulty here lies in that the walker state at time $t_0$ can now depend on $\{A(t)\}_0^{t_0}$, something that was not possible when all configurations were connected.
However, $\bpsi$ is easily characterized when the underlying transitions between configurations are described by a Markov chain and the times that the graph remain in a configuration correspond to mutually independent sequences of iid random variables; one sequence for each configuration.  
Let $\bP = [p_{ij}]$ denote the $m\times m$ transition probability matrix for the graph configurations at the time of transitions between graphs and, with an abuse of notation, let $\{S_{k,i}\}_0^\infty$,  $k=1,\ldots m$, denote the mutually independent iid sequences of configuration holding times for the graph configurations.   
 
We focus now on transitions that the walker makes between connected components in two different graph configurations, say the $j_1$-th connected component in configuration $G_{k_1}$  and the $j_2$-th connected component in configuration $G_{k_2}$.  
We define a transition probability matrix $\hat{\bP} = [\hat{p}_{k_1,j_1;k_2,j_2}]$ as follows
\beq \label{eq:P-disc} 
\hat{p}_{k_1,j_1;k_2,j_2} = p_{k_1,k_2} \frac{\sum_{v\in V_{k_1,j_1}\cap V_{k_2,j_2}} d_{k_1,v}/\beta_{k_1,v}}{\sum_{w\in V _{k_1,j_1}} d_{k_1,w}/\beta_{k_1,w} }  .
\eeq
The first term accounts for transitions between graph configurations and the second term accounts for the walker dynamics. Here 
$\hat{\bP}$ can be thought of as the transition probability matrix for a discrete time Markov chain that characterizes the subgraphs visited by a random walk at graph transitions in the limit as $\gamma\rightarrow \infty$.  
This chain is irreducible provided the graph is T-connected.   Let  $\bpsi^* = (\psi^*_{1,1},\ldots, \psi^*_{m,l_m})$ denote the stationary distribution of this MC. The earlier introduced probability distribution $\bpsi$ can be expressed in terms of $\bpsi^*$ as follows
\begin{equation}
\label{eq:RW-disc}
\psi_{k,l} = \frac {\psi^*_{k,l} E[S_k] } {\sum_{i=1}^m \sum_{j=1}^{o_i} \psi^*_{i,j} E[S_i]}, \quad A_k \in \cA;l=1,\ldots,o_k \, ,
\end{equation}
Note that the above characterization depends on the independence and identical distribution assumptions of the configuration holding times.
This, along with (\ref{eq:theor-fast-gamma-disc}) fully characterizes the stationary distribution of the walker in the fast walker regime for the Markovian environment.
}

%% file: slow_alt-new.tex
\subsubsection{The Slow Walker}\label{sec:slow_walker}

In this section we consider the walker stationary distribution, $\bpi$, at the other timescale decomposition, namely where the graph dynamics speed up relative to the walker.  Consider a walker with the set of walker rates $\Gamma$ \linebreak walking  a dynamic graph $\Psi$.  We consider the RMPP $\Psi^{(a)}=\{(X_i,a S_i)\}_{i \in \mathbb{Z}}$ that is a speed up of the RMPP $\Psi$ by a factor of $a$, $0 < a < 1$, and characterize the CTRW on  $\Psi^{(a)}$ as $a\rightarrow 0$.    
We denote by $A^{(a)}(t)$ the state at time $t$ of the dynamic graph corresponding to the RMMP $\Psi^{(a)}$. We do this in two steps.  We first consider an observer of $\Psi^{(a)}$, who makes observations according to a renewal process with the property that it has a continuous non-increasing probability density function with finite mean.
We then determine conditions under which the observer is guaranteed to observe independent instances of the graph with probability given by the stationary distribution of the graph.  Finally, we consider a Poisson observer and couple the walker with it in order to characterize the stationary distribution of the walker.

We introduce a renewal process $\{W_j\}_1^\infty$ where $W_j$ denotes the time between the $(j-1)$-th and $j$-th observations with CDF $G(x)$ (with PDF $g(x)$) satisfying the following assumption.

\begin{ass} \label{ass:pdf}
The pdf $g(x) := dP(W_i<x)/dx$  is differentiable
with $g'(x) = dg(x)/dx$, non-increasing, nonnegative, 
with (i) $g(0)<\infty$, (ii) $\int_0^\infty g(x)dx = 1$ and (iii)  $E[W_j]=\int_0^\infty x g(x) dx:=D<\infty$.
As a consequence of the previous assumptions (iv) $\int_0^\infty x g'(x) dx = -1$
(Hint: use an integration by part and note that $\lim_{x\to \infty} xg(x)=0$ thanks to (iii)).

\end{ass}
Assumption \ref{ass:pdf} holds  if $W_i$ is exponentially distributed with parameter $\gamma < \infty$. It also holds
if $W_i$ has a Pareto distribution with Pareto index strictly larger than one. 
We will now observe the graph at the renewal instants  $\sum_{k=1}^j W_k$, $j\geq 1$. 

We denote by $A^{(a)}(t)$ the state at time $t \in \mathbb{R}$ of the graph  associated with the RMPP $\Psi^{(a)}$, namely,
\begin{equation}
A^{(a)} (t)=\sum_{i=-\infty}^\infty X_i {\bf 1}(aT_i \leq t <aT_{i+1}),
\label{def-Aa}
\end{equation}
so that $A^{(a)}(t)=A(t/a)$ for all $t\in \mathbb{R}$, where $T_i$ is as defined in~\eqref{eq:At}. We denote by $A^{(a)}_{(j)}=A^{(a)}(W_1+\cdots+W_j)$ the graph 
configuration of $A^{(a)}(t)$ at the $j$-th renewal instant ($j\geq 1$).

\begin{lem}
Let $\Psi^{(a)}$ be the RMPP associated with the stationary and ergodic dynamic graph $\Psi$.
If the observation process satisfies Assumption \ref{ass:pdf}, then for any $i\geq 1$, $k=1,\ldots,m$,
\begin{equation}
\lim_{a \to 0} P\left(A^{(a)}_{(i+1)}=A_k\,|\, A^{(a)}_{(j)}=A_{l_j}, j=1,\ldots,i\right)=\sigma_k.
\label{sw:thm4-3}
\end{equation}
\end{lem}

\begin{pf} 
Throughout $i\geq 1$ is fixed and so are $k,l_1,\ldots, l_i \in \{1,\ldots,m\}$. Define the set $I_i=\{1,\ldots,i\}$ and
let  $g_a(x):=a g(ax)$.

Conditioning on $W_j=y_j$ for $j\in I_i$ and $W_{i+1}=x$ and using the independence assumption between the process
$\{A(t)\}$ and the iid rvs $(W_j)_j$  with pdf $g(\cdot)$,  gives
\begin{eqnarray}
\lefteqn{ P\left( A_{(i+1)}^{(a)}=A_k\,|\, A_{(j)}^{(a)}=A_{l_j},j\in I_i\right)-\sigma_k}\nonumber \\
&=&
\int_{[0,\infty)^i} \int_{x=0}^\infty  (P(A((x_i +x)/a ) =A_k\,|\,  A(x_j/a) =A_{l_j}, j\in I_i)-\sigma_k) g(x) dx   \prod_{j\in I_i} g(y_j) dy_j\nonumber\\
&=&\int_{[0,\infty)^i} \int_{x=0}^\infty  (P(A(x_i +x ) =A_k\,|\, A(x_j) =A_{l_j}, j\in I_i) -\sigma_k) g_a(x) dx   \prod_{j\in I_i} g_a(y_j) dy_j\nonumber\\
&=&\int_{[0,\infty)^i} \int_{x=0}^\infty  (P(A(x ) =A_k\,|\, A(x_j-x_i) =A_{l_j}, j\in I_i) -\sigma_k) g_a(x) dx   \prod_{j\in I_i} g_a(y_j) dy_j
\label{int200}
\end{eqnarray}
with $x_j:=\sum_{l=1}^j y_j$,  $j\in I_i$, where we have used the stationarity of the process $\{A(t)\}_{-\infty}^\infty$ to derive
(\ref{int200}).
Define 
\begin{align*}
{\bf x}^i&:=(x_1,\ldots,x_i),\,\\
f(u,{\bf x}^i)&:= P(A(u) =A_k\,|\, A(x_j-x_i) =A_{l_j}, j\in I_i)) -\sigma_k \, ,\\
F(x,{\bf x}^i)&:=\int_0^x f(u,{\bf x}^i)du. 
\end{align*}
Note that $|f(u,{\bf x}^i)|\leq 1$ for  any
$u, {\bf x}^i$ so that $|F(x,{\bf x}^i)|\leq x$ for any $x,{\bf x}^i$. In this notation (\ref{int200}) rewrites
\begin{eqnarray}
P\left(A_{(i+1)}^{(a)}=A_k\,|\, A_{(j)}^{(a)}=A_{l_j},j\in I_i\right)-\sigma_k=\int_{[0,\infty)^i} \int_{x=0}^\infty f(x,{\bf x}^i)g_a(x) dx   \prod_{j\in I_i} g_a(y_j) dy_j.
\label{int220}
\end{eqnarray}
Integrating by parts and using the definition of $g_a(x)$ and Assumption~\ref{ass:pdf} yields 
\begin{eqnarray}
\int_0^\infty f(s,{\bf x}^i) g_a(x)dx &= & \left[g_a(x) F(x,{\bf x}^i)\right]_{0}^\infty -  \int_0^\infty  F(x,{\bf x}^i) g'_a(x)dx
\nonumber\\
&=& \lim_{x\uparrow \infty} g_a(x)F(x,{\bf x}^i) -  a^2 \int_0^\infty  F(x,{\bf x}^i) g'(ax)dx
\label{int2000}\\
&=&- a^2\int_0^\infty F(x,{\bf x}^i) g'(ax) dx \, ,
\label{int201}
\end{eqnarray}
where the limit in (\ref{int2000}) is zero since $0\leq g_a(x)F(x,{\bf x}^i) \leq g_a(x) x$ and 
$\lim_{x\to\infty} x g(x)=0$ (see Assumption \ref{ass:pdf}). Combining (\ref{int220}) and 
(\ref{int201}) gives
\begin{eqnarray}
P\left (A_{(i+1)}^{(a)}=A_k\,|\, A_{(j)}^{(a)}=A_{l_j},j\in I_i\right)-\sigma_k= -a^2\int_{[0,\infty)^i} \int_{x=0}^\infty F(x,{\bf x}^i) g'(ax) dx \prod_{j\in I_i} g_a(y_j) dy_j.
\label{int221}
\end{eqnarray}
Fix $\epsilon>0$. By Prop. \ref{prop:uniform} we know that there exists $0<T_{\epsilon}<\infty$,
denoted as $T$ from now on, such that for all $x > T$, $|F(x,{\bf x}^i)/x|< \epsilon/2 $,
uniformly in ${\bf x}^i$ or, equivalently, uniformly in $y_1,\ldots,y_i\in [0,\infty)$.
We have (Hint: use Assumption \ref{ass:pdf} and inequality
$|F(x,{\bf x}^i)|\leq x$)
\begin{eqnarray}
\left |\int_0^\infty F(x,{\bf x}^i) g'(ax) dx \right |  & \leq &  - \int_{0}^T F(x,{\bf x}^i) g'(ax) dx - \int_{T}^\infty \left |\frac{F(x,{\bf x}^i)}{x} \right |x g'(ax) dx\nonumber\\
&\leq & -T \int_{0}^T g'(ax) dx  -\frac{\epsilon}{2}  \int_{T}^\infty x g'(ax) dx\nonumber\\
&\leq&  -T \int_{0}^\infty g'(ax) dx  - \frac{\epsilon}{2}  \int_{0}^\infty x g'(ax) dx\nonumber\\
&=& \frac{T g(0)}{a} +\frac{\epsilon}{2a^2}
\end{eqnarray}
so that, from (\ref{int221}),
\begin{eqnarray}
 \left | P\left( A_{(i+1)}^{(a)}=A_k\,|\, A_{(j)}^{(a)}=A_{l_j},j\in I_i\right)-\sigma_k\right| &\leq&\left(a T g(0) +\frac{\epsilon}{2}\right) \int_{[0,\infty)^i} \prod_{j\in I_i} g_a(y_j) dy_j\nonumber\\
&=&a T g(0) +\frac{\epsilon}{2}.
\label{int203}
\end{eqnarray}
We observe from (\ref{int203})   that 
\[
\left  | P\left(A_{(i+1)}^{(a)}=A_k\,|\, A_{(j)}^{(a)}=A_{l_j},j\in I_i\right )-\sigma_k\right|  <\epsilon
 \]
 for  any $0<a<\epsilon/(2 T g(0))$, 
 which completes the proof since $\epsilon$ is arbitrary.
\end{pf}

Application of the chain rule yields the following result.
\begin{prop} \label{propslow}
As $a\to 0$ the sequence $\{A_{(j)}^{(a)}\}_{j \geq 1}$ converges to an iid sequence with distribution 
\linebreak
$P\left(A^{(a)}_{(j)} = A_k\right) = \sigma_k$, $k=1,\ldots , m$.
\end{prop}

It is now straightforward to describe the behavior of  a constant rate walker.  Take the observation process to be Poisson with rate $\gamma^\prime > \gamma$ (recall that $\gamma$ is the walker rate of our constant rate walker), namely $P(W_j< x)=1-e^{-\gamma^\prime x}$.
Consider the dynamic graph $\{A^{(a)}(t)\}_{t \in \mathbb{R}}$ associated with the RMPP $\Psi^{(a)}$  (see (\ref{def-Aa})).
We embed the times that the walker takes a step into the observation process.  At each observation the walker takes a step with probability $\gamma/\gamma^\prime$; otherwise it does not with probability $(\gamma^\prime - \gamma)/\gamma^\prime$.  Let $U^{(a)}_j\in \{1,\ldots,n\}$ denote the position of the walker immediately after the $j$-th observation of the observation process ($j\geq 1$), and let  $U^{(a)}_{0}$ be the walker position at time $t=0$.
We assume that $\lim_{a\to 0}P(U^{(a)}_{0}=v)$ exists for all $v=1,\ldots,n$. This is the case, for instance, if $U^{(a)}_0$ is constant for any $a>0$.

The following equation describes the behavior of $\pi^{(0)}_j(v) = \lim_{a\to 0}P(U^{(a)}_j = v)$, the fraction of {\em observations} after which the walker resides at node $v \in V$. Assume first that this limit
exists for any $j\geq 1$ and $v=1,\ldots,n$. 
Let  $\bP := [\bP(u,v)]$ be an $n$-by-$n$ stochastic matrix with $(u,v)$-entry given by
\begin{equation} 
\bP(u,v) = \begin{cases}
               \frac{\gamma}{\gamma^\prime} \sum_{k=1}^m \sigma_k  \frac{A_{k}(u,v)}{d_{k,u} } \,,& \mbox{if } d_{k,u} > 0, \, u \neq v, \\
                \frac{\gamma^\prime - \gamma}{\gamma^\prime}  + \frac{\gamma}{\gamma^\prime}\sum_{k=1}^m  \sigma_k \bone_{\{d_{k,u} = 0 \}} \, , & \mbox{if }   u = v \, , \\
                0\,, & \mbox{otherwise,}
                \end{cases}
\label{eq:P}
\end{equation}
$u,v = 1,\ldots,n$.
We have
\begin{align}
 \pi^{(0)}_j(v) &=\lim_{a \to 0}\sum_{u=1}^n  \sum_{k=1}^mP\left(A^{(a)}_{(j)}= A_k, U^{(a)}_{j-1} = v\right)  \times \sigma_k \left(\frac{A_{k}(u,v)}{d_{k,u} }{\bf 1}(d_{k,u}>0)+ {\bf 1}(u=v){\bf  1}(d_{k,u}=0)\right) \nonumber\\
 & = \sum_{u=1}^n \sum_{k=1}^m  \lim_{a \to 0} P\left(A^{(a)}_{(j)} = A_k\right)  \lim_{a\to 0}P\left(U^{(a)}_{j-1} = v\right)
 \sigma_k \left(\frac{A_{k}(u,v)}{d_{k,u} }{\bf 1}(d_{k,u}>0)+ {\bf 1}(u=v){\bf  1}(d_{k,u}=0)\right) \nonumber\\
 &=\sum_{u=1}^n \bP(u,v) \pi^{(0)}_{j-1}(u), 
 \label{int400}
\end{align}
for $j\geq 1$ and $v=1,\ldots,n$, where $d_{k,u}$ is the degree of node $u\in V$ in graph configuration $A_k$ and $A_k(u,v)$ is the $(u,v)$-entry of the adjacency matrix $A_k$ (i.e. $A_k(u,v)=1$ if there is a link between vertices $u$ and $v$ in configuration $k$ and zero otherwise).
The second and third equalities follow from Proposition \ref{propslow}.

The existence of $\lim_{a\to 0}P(U^{(a)}_{j}=v)$ for all $j\geq 1$ and $v=1,\ldots,n$ can be shown by induction on $j$ based on (\ref{int400}).

Define  $\pi^{(0)}_j:= (\pi^{(0)}_j(1),\ldots, \pi^{(0)}_j(n))$.
With these definitions (\ref{int400}) rewrites in the following matrix form
\[
\pi^{(0)}_j = \pi^{(0)}_{j-1} \bP , \quad j\geq 1,
\]
with $\pi^{(0)}_0$ the  $n$-dimensional vector where all entries are equal to zero except entry $U^{(a)}_{0}$  that is equal to $1$.
To show that $\pi^{(0)}$ is unique, we prove that $\bP$ is irreducible.
The definition of T-connectivity and the fact that $\sigma_k > 0$, $k=1,\ldots,m$ 
imply that  $\bP$ is an adjacency matrix of a strongly connected directed graph and 
strong connectivity of this graph is equivalent to the irreducibility of $\bP$~\cite[Theorem 6.2.24]{Horn}, 
the details of this proof are found at Appendix~\ref{Pirred}. 
As $\bP$ is also aperiodic, as the diagonal elements of $\bP$ are non-zero, 
then the associated discrete-time Markov chain is ergodic since the
state-space is finite so that,  by Markov chain theory, the limit 
$\pi^{(0)}=\lim_{j\to\infty}\pi^{(0)}_j$ exists and is given by the unique solution of 
\[ 
\pi^{(0)} = \pi^{(0)} \bP,  \quad \sum_{u=1}^n \pi^{(0)}(u)=1.
\]
Because of the PASTA property the steady state probability $\pi^{(0)}=\lim_{j\to\infty}\pi^{(0)}_j$ on the observation events is also the distribution in time, i.e., in the limit as $a \to 0$, $\bpi = \pi^{(0)}$.

The case where the walker rate depends on the node and graph configuration in which the walker resides yields a similar characterization.  

\begin{prop} \label{cor:joint_slow_walker}
Let $\Psi$ be a stationary, ergodic, and T-connected dynamic graph and let $\{(A(t),U(t))\}_{t\geq 0}$ be an associated CTRW with walker rates $\Gamma = \{\beta_{k,v} : k=1,\dots,m,\,v \in V\}$.
Let $\beta_{\max} > \sup(\Gamma)$.
Then, in the limit as $a \to 0$, the stationary walker position distribution $\bpi$ satisfies
\[ \bpi = \bpi \bP \, , \]
where 
\begin{equation} \label{slowMC}
\bP(u,v) = 
\begin{cases} 
\sum_k \sigma_k \frac{\beta_{k,v}}{\beta_{\max}} \frac{A_{k,vu}}{d_{k,v}}, & d_{k,u} > 0, \\
&  \quad v\ne u \\
\sum_k \sigma_k \bigl(1- \frac{\beta_{k,v}}{\beta_{\max}} + \frac{\beta_{k,v}}{\beta_{\max}}\bone_{\{d_{k,u} = 0 \}}\bigr), & v=u \, .
\end{cases}
\end{equation}
\end{prop}
The proof is similar to that given for the constant rate walker case and is omitted.  The assumption that $\beta_{\max} > \sup(\Gamma)$ ensures that $\bP$ is aperiodic and the assumption of T-connectivity ensures that $\bP$ is irreducible.

Proposition~\ref{cor:joint_slow_walker} shows, as $a \to 0$, the steady state distribution of a walker with state dependent rates to be just a function of $\bsigma$ (the stationary distribution of $\{A(t)\}_{t \geq 0}$), the set of configurations $\cA$, and the walker rates $\vbeta$, regardless of the graph dynamics.

%% file: stable_walker.tex
\subsection{Time-scale Invariant Stationary \\ Distribution}\label{sec:time_scale_inv}
In this section we turn our attention to a sufficient condition where the CTRW stationary distribution is invariant to the walker time scale $\gamma$.
Consider a CTRW $\{(A(t),U(t))\}_{t \geq 0}$ with non-zero walker rates on a stationary, ergodic, and T-connected graph $\Psi$.

The key insight into our sufficient condition is the following:
If there exists a $\bpi$ that is the CTRW stationary distribution given any (static) configuration $A_1,\ldots, A_m$, then once the CTRW reaches distribution $\bpi$ it remains with distribution $\bpi$ independent of the graph dynamics.
The key challenge is to show that the CTRW always converges to distribution $\bpi$, irrespective of the graph dynamics.
We see that this is true if $\Psi$ is stationary, ergodic, and T-connected.
However, we also believe that the following results can be extended to some families of non-stationary dynamic graphs.

We first present the notation used in this section.
It will be useful to describe the walker rates as a row vector $\vgamma(\gamma) = (\gamma \beta_{k,v})_{v \in V, k=1,\dots,m}$.

The CTRW confined to a given configuration $A_k$ is a Markov chain, $k=1,\ldots,m$.
Let 
\[
Q(A_k,\vgamma(\gamma)) = \begin{cases}
     A_k(i,j) \gamma \beta_{k,i}/d_{k,i} \, , & \mbox{if } i\neq j \, ,\\
     -\sum_{j \in V} A_k(i,j) \gamma \beta_{k,i}/d_{k,i} \, , & \mbox{if } i = j \, ,
\end{cases}
\]
where $A_k(i,j)$ is the element $(i,j)$ of $A_k$,
be the infinitesimal generator of $\{U(t)\}_{t \geq 0}$ given configuration $A_k$.

\begin{ass} [Fixed point $\bpi^\star$] \label{ass:bpi}
Let $\cA$ be the set of graph configurations of $\Psi$ and $\vgamma$ be a set of walker rates such that there exists a $\bpi^\star \in [0,1]^n,$ $\sum_{v \in V} \bpi^\star(v) = 1$, 
that is a fixed point solution to
\begin{equation} \label{eq:fixed}
 {\bf 0} = \bpi^\star Q(M,\vgamma) \: , \quad \forall M \in \cA \, .
\end{equation}
\end{ass}

In what follows we show that if $\cA$ and $\vgamma$ satisfy Assumption~\ref{ass:bpi}, then $\lim_{t \to \infty} P[U(t) = v] = \bpi^\star(v)$, $\forall v \in V$. 
Moreover, we show that $\bpi^\star$ is independent of $\gamma$.
We are now ready for the main result of this section.
\begin{theo} \label{thm:samefixed}
Let graph process $\Psi$ be a stationary, ergodic,  and T-connected graph with configuration set $\cA$. 
Let  $\{(A(t),U(t))\}_{t \geq 0}$ be a CTRW  on the dynamic graph $\{A(t)\}_{t \in \mathbb{R}}$ associated with $\Psi$.
The CTRW has walker rates $\vgamma(\gamma) = \{\gamma \beta_{k,v}\}, v \in V, k=1,\dots,m$, $\gamma > 0$.
If $\cA$ and $\vgamma(1)$ satisfy Assumption~\ref{ass:bpi}, then $$\lim_{t \to \infty} P[U(t) = v] = \bpi^\star(v)\,, \quad \forall v \in V,$$
where $\bpi^\star$ solves~\eqref{eq:fixed}.
Moreover, $\bpi^\star$ does not depend on $\gamma$.
\end{theo}
\begin{pf}
For now assume $\gamma=1$.
Let $\Pi(t) = (P[U(t)=v])_{v \in V}$.
The Kolmogorov forward equation gives
\begin{equation}\label{eq:Kol}
\frac{d \Pi(t)}{dt} = \Pi(t) Q(A(t),\vgamma(1)).
\end{equation}
From Assumption~\ref{ass:bpi} there exists $\bpi^\star$ is that is a solution to~\eqref{eq:fixed}.
Hence, $\Pi(t) = \bpi^\star$ is also a solution to~\eqref{eq:Kol} where ${d \Pi(t)}/{dt}=0$.
It follows from Theorem \ref{CTRW-stionary} that there is no other solution to~\eqref{eq:Kol} and therefore  $\lim_{t \to \infty} P[U(t) = v] = \bpi^\star(v)$, $\forall v \in V$.
To show that $\bpi^\star$ does not depend on $\gamma$, note that for any $\alpha > 0$ and $k=1,\ldots, m$
\begin{align*}
{\bf 0}\alpha = \bpi^\star Q(A_k,\vgamma(1))\alpha =  \bpi^\star Q(A_k,\vgamma(\alpha)).
\end{align*}
\end{pf}

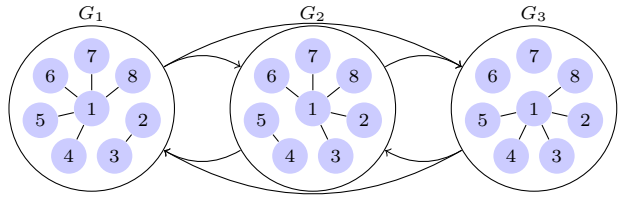
\begin{figure}[h]
\begin{center}
\begin{tikzpicture}[scale=.7,auto=left,every node/.style={circle,fill=blue!20}]
{
\scriptsize
\node [fill=none,draw,circle,minimum size=2.2cm] (s1) at (0,0) {};

\domath{0}{\offset}

\node  (a1) at (\offset+0,0) {1};

\foreach \n in {2,...,8}{%
\domath{\offset+1*sin(\n*360/7)}{\x}
\domath{1*cos(\n*360/7)}{\y}
\node (a\n) at (\x,\y) {\n};
};

\foreach \n in {4,...,8}
\draw (a1) -- (a\n);

\draw (a2) -- (a3);

\node [fill=none] (n1) at (\offset,1.8) {$G_1$};

\domath{4.2}{\offset}

\node [fill=none,draw,circle,minimum size=2.2cm] (s2) at (\offset,0) {};

\node  (b1) at (\offset+0,0) {1};

\foreach \n in {2,...,8}{%
\domath{\offset+1*sin(\n*360/7)}{\x}
\domath{1*cos(\n*360/7)}{\y}
\node (b\n) at (\x,\y) {\n};
};

\node [fill=none] (n2) at (\offset,1.8) {$G_2$};

\foreach \n in {2,3,6,7,8}{%
\draw (b1) -- (b\n);
};

\draw (b4) -- (b5);

\path[->] (s1) edge [bend left] (s2);
\path[->] (s2) edge [bend left] (s1);

\domath{\offset + 4.2}{\offset}

\node [fill=none,draw,circle,minimum size=2.2cm] (s3) at (\offset,0) {};

\node  (c1) at (\offset+0,0) {1};

\foreach \n in {2,...,8}{%
\domath{\offset+1*sin(\n*360/7)}{\x}
\domath{1*cos(\n*360/7)}{\y}
\node (c\n) at (\x,\y) {\n};
};

\node [fill=none] (n3) at (\offset,1.8) {$G_3$};

\foreach \n in {2,3,4,5,8}{%
\draw (c1) -- (c\n);
};


\path[->] (s1) edge [bend left] (s3);
\path[->] (s2) edge [bend left] (s3);

\path[->] (s3) edge [bend left] (s1);
\path[->] (s3) edge [bend left] (s2);
}
\end{tikzpicture}
\end{center}
\vspace{-10pt}
\caption{\small Illustration of a dynamic graph whose node degrees are either kept constant in all graph configurations or there are isolated nodes. We make no assumption about the holding times of each graph configuration except that the graph is T-connected.\label{fig:dyngraph1}}
\end{figure}

Examples of adjacency matrix sets $\cA = \{A_k : k=1,\ldots,m\}$ that satisfy Assumption~\ref{ass:bpi} for a constant rate walker, $\vgamma(\gamma)=(\gamma,\ldots, \gamma)$ include:
\begin{itemize}
\item {\bf Regular graphs}:
$A_i$, $i=1,\ldots,m$, consists of $C_i \ge 1$ connected components where the $j$-th 
connected component ($j=1, \ldots, C_i$) is a $d_j^{(i)}$-regular graph ($d_j^{(i)} >= 0$).
\item {\bf Nodes $v \in V$ alternate between isolated and connected with constant degree, i.e., $d_k(v) \in \{0, d(v)\}$, $d(v) > 0$, $k = 1,\ldots,m$.} Figure~\ref{fig:dyngraph1} illustrates a dynamic graph that satisfies these requirements.
\end{itemize}

Conditions imposed on the walker rates $\vgamma(\gamma)$ can also guarantee that Assumption~\ref{ass:bpi} is valid for any set of graph configurations $\cA$, as long as $\{A(t)\}_{t \in \mathbb{R}}$ is T-connected.
Consider the coupled CTRW and graph dynamics that satisfies Assumption~\ref{ass:bpi} in the following proposition, stated without a proof: 
\begin{prop}[Degree proportional walker]  \label{prop:deg} 
Let $\cA = \{A_k : k=1,\ldots,m\}$ be a set of graph configurations. If the walker rates are $\vgamma(\gamma) = (\gamma d_{k,v})_{v \in V, k=1,\dots,m}$, where $d_{k,v}$ is the degree of node $v$ at configuration $k$. Then Theorem~\ref{thm:samefixed} is satisfied. Moreover, $\bpi^\star = (\frac{1}{n},\ldots,\frac{1}{n})$.
\end{prop}

Proposition~\ref{prop:deg} has an interesting application. We can uniformly sample nodes (in time) without knowing the underlying topologies, $\mathcal{A}$, or graph dynamics, $\{A(t)\}_{t \in \mathbb{R}}$, as long as $\{A(t)\}_{t \in \mathbb{R}}$ is stationary, ergodic, and T-connected.
So far we have focused on conditions that allow us to obtain the stationary distribution of the walker.
In what follows we present some case studies solved numerically.

%% file: casestudies.tex
\section{Case Studies}\label{sec:casestudies}

\input{markov}
\input{star_circle}

\input{edge_markovian}

\input{buslines_example}


%% file: markov.tex
In previous sections we characterized the stationary behavior of an RW on
a broad class of dynamic graph processes $\{(X_i,S_i)\}_{i \geq \mathbb{Z}}$. 
In this section, we focus mostly on random walks on {\em Markovian dynamic 
graphs} where $\{X_i\}_{i \geq 0}$ forms a Markov chain and $\{S_i\}_{i \geq 0}$ is a sequence of independent and exponentially 
distributed random variables. Note that this model allows state dependent graph holding times to be taken into account 
in the Markov chain $\{X_i\}_{i \geq 0}$.
In what follows we provide several examples and numerical evaluations. Later in the section, we also show how the proposed framework 
can be applied to a Delay-Tolerant Network (DTN) scenario.


\subsection{Markovian dynamic graph examples}
In this section we present numerical results for some toy Markovian dynamic graph examples, that illustrate the 
non-trivial behavior of random walks on dynamic graphs and that support our theoretical findings.

%% file: star_circle.tex
\subsubsection{Star-circle example}

\begin{figure*}[ht]
\centering
\subfloat[\label{fig:star-circle-a} The star-circle graph dynamics.]{
\includegraphics[width=3in]{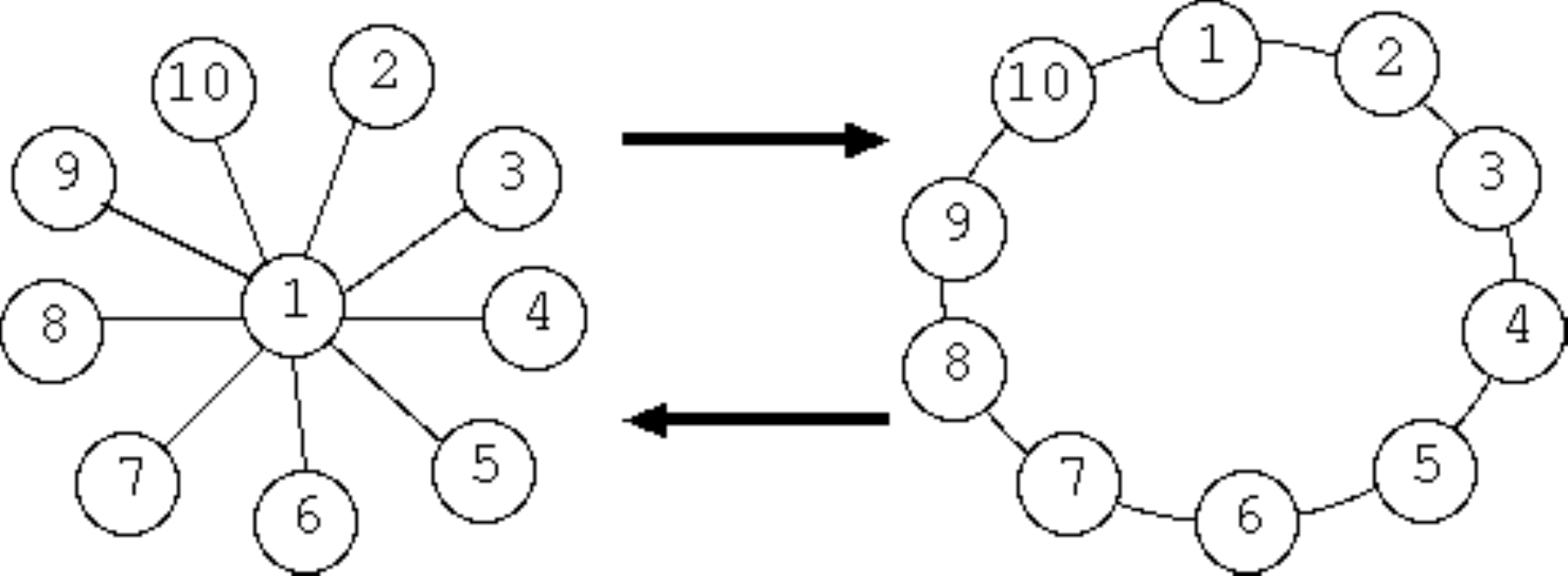}}
\hspace{1cm}
\subfloat[\label{fig:star-circle-b} Steady state distribution of walker as a function of walker rate (nodes 5, 6 ,7 not shown for clarity).]{
\includegraphics[width=3.4in]{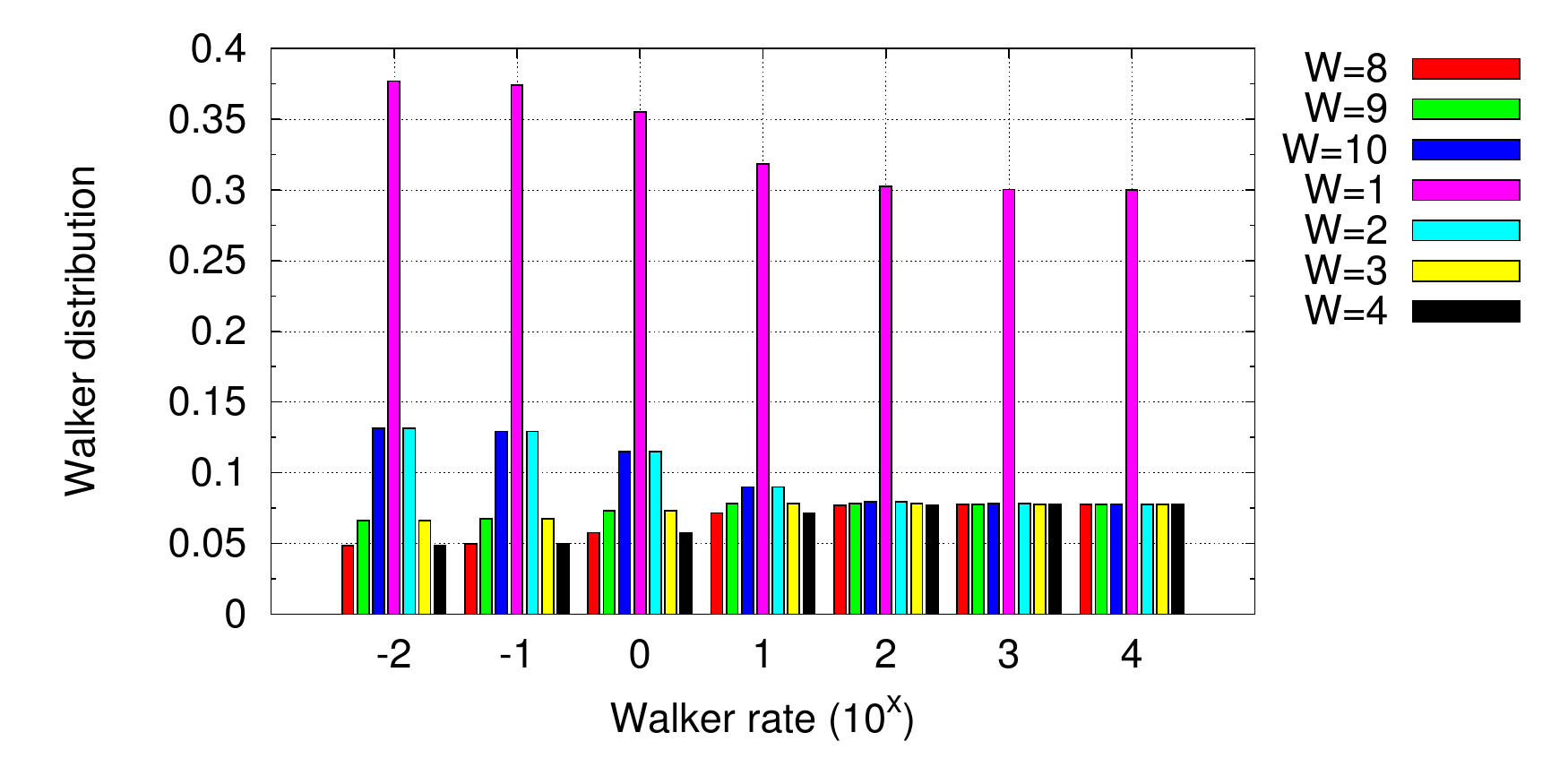}}
\caption{\label{fig:star-circle} The star-circle graph dynamics and behavior of CTRW as a function of walker rate.}
\end{figure*}

We begin by considering a very simple model, consisting of just two graph snapshots: a star and a circle, as illustrated in Figure~\ref{fig:star-circle-a}. The graph transits from one snapshot to another with rates $\lambda_{12} = \lambda_{21} = 1$. Thus, the average time in each 
graph is 1/2. Note however, that edges (1,2) and (1,10) are always present, since they exist in both configurations.

We investigate the steady state solution of the CTRW on this dynamic graph. 
Figure~\ref{fig:star-circle-b} shows the steady state distribution of the random walk as a function of the walker 
rate (for clarity, only a subset of the states are shown). Note that the stationary distribution of the walker 
depends on the walker rate and converges to different distributions as the walker moves faster or slower. 
Indeed, the numerical results obtained are in agreement with the theoretical distributions for 
the fast and slow walker given in Sections~\ref{sec:fast} and~\ref{sec:slow_walker}, respectively. Finally, 
we note that in a graph with $n$ nodes, node one alternates between degrees $n-1$ and two.
As $n$ increases the dependence on the walker speed is magnified.

%% file: edge_markovian.tex
\subsection{Edge Markovian model and examples}\label{sec:edge_markovian}

In this section we consider some particularities of random walks on a special class of dynamic graphs called edge 
Markovian graphs.
Start with a fixed adjacency matrix $A$ and attach an independent On-Off process to each edge with exponentially distributed holding times.
In edge Markovian graphs, edges alternate between being present and absent from the graph according to independent On-Off processes. 
Let $E$ be the set of edges in the graph described by $A$.
Let $\Lambda_0(e)$ and $\Lambda_1(e)$ denote the rate at which edge $e \in E$ changes from 
the On to the Off state and from the Off to the On state, respectively, which can 
vary from edge to edge. 

A few observations on the model follows. Let $A_k$ be a particular configuration
of the dynamic graph model.
In particular, the edge Markovian model induces a total of $m = 2^{|E|}$ configurations, 
which represent all possible labeled subgraphs over an edge set with $|E|$ edges. Moreover, 
consider any transition between two configurations induced by the model. The graphs 
corresponding to these two configurations differ by exactly one edge, since the on-off 
processes associated with the edges are continuous in time. Moreover, the rate associated 
with this graph transition is given by the corresponding rate of the edge (either its On 
rate or Off rate).

Let $q_{uv}$ denote the stationary fraction of time that edge $(u,v) \in E$ is On, which is simply 
given by $$q_{uv} = \Lambda_1(u,v)/(\Lambda_0(u,v)+\Lambda_1(u,v)).$$ Let $\sigma_k$ denote the 
steady state fraction of time that the edge Markovian model spends in configuration 
$A_k$, $k=1,\ldots,2^{|E|}$. 
In particular, we have  
\begin{equation}
\begin{split}
\sigma_k = & \prod_{(u,v) \in E} {\bf 1}\{A_k(u,v)\} q_{uv} + (1 - {\bf 1}\{A_k(u,v)\})(1 - q_{uv}) 
\label{eq:PG_k}
\end{split}
\end{equation}
Note that $\sigma_k$ is given by the product of the probabilities the egdes that define $A_k$ are present while all other edges are absent. Since 
all edges are independent, this is trivially obtained as shown above.
Edge Markovian graphs are of interest due to their simple description and structure. However, as we will soon see in our numerical results, the steady state distribution of a constant rate random walk on this model depends on the walker rate. 
It is an open problem whether the walker steady state distribution can be obtained in closed form.

\subsubsection{Edge Markovian: $n=3$}

Consider the edge Markovian graph model over a complete graph with 3 nodes. 
The number 
of different configurations is $2^3 = 8$, out of which 4 have at least one isolated vertex (graph 
not connected). Moreover, let $\Lambda_1(1,2) = \Lambda_0(1,2) = 10^4$, $\Lambda_1(1,3) = \Lambda_0(1,3) = 1$, $\Lambda_1(2,3) = \Lambda_0(2,3) = 1$. Thus, $p_e = 0.5$ for every edge (i.e., all edges have the same time average), and thus, all configurations have the same time average 
probability of 1/8, as given by equation (\ref{eq:PG_k}). 

\begin{figure*}[ht]
\centering
\subfloat[\label{fig:K3-a} Steady state distribution of walker as a function of walker rate.]{
\includegraphics[width=2.8in,height=2in]{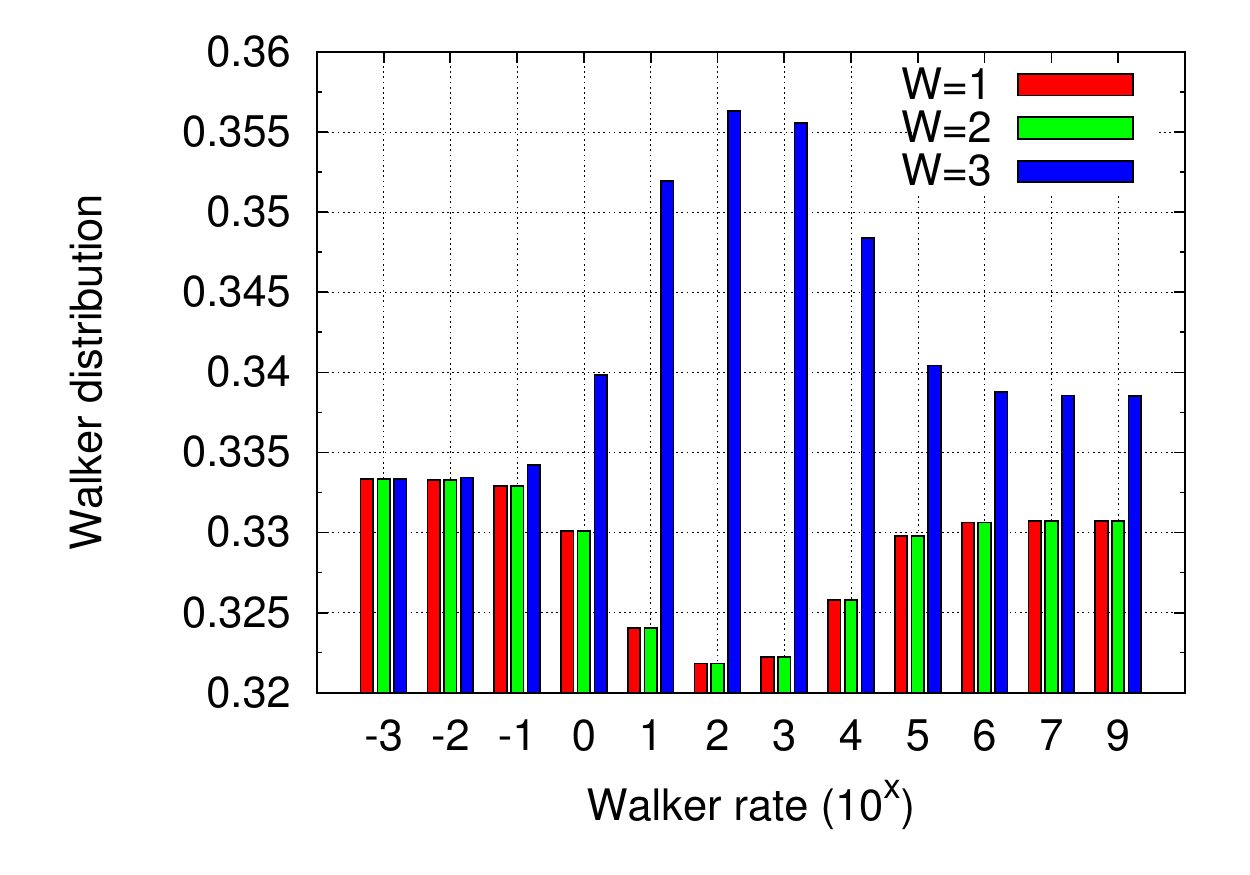}}
\hspace{1cm}
\subfloat[\label{fig:K3-b} Total variation distance between theoretical results for fast and slow walker when compared to exact walker distribution.]{
\includegraphics[width=2.8in,height=2in]{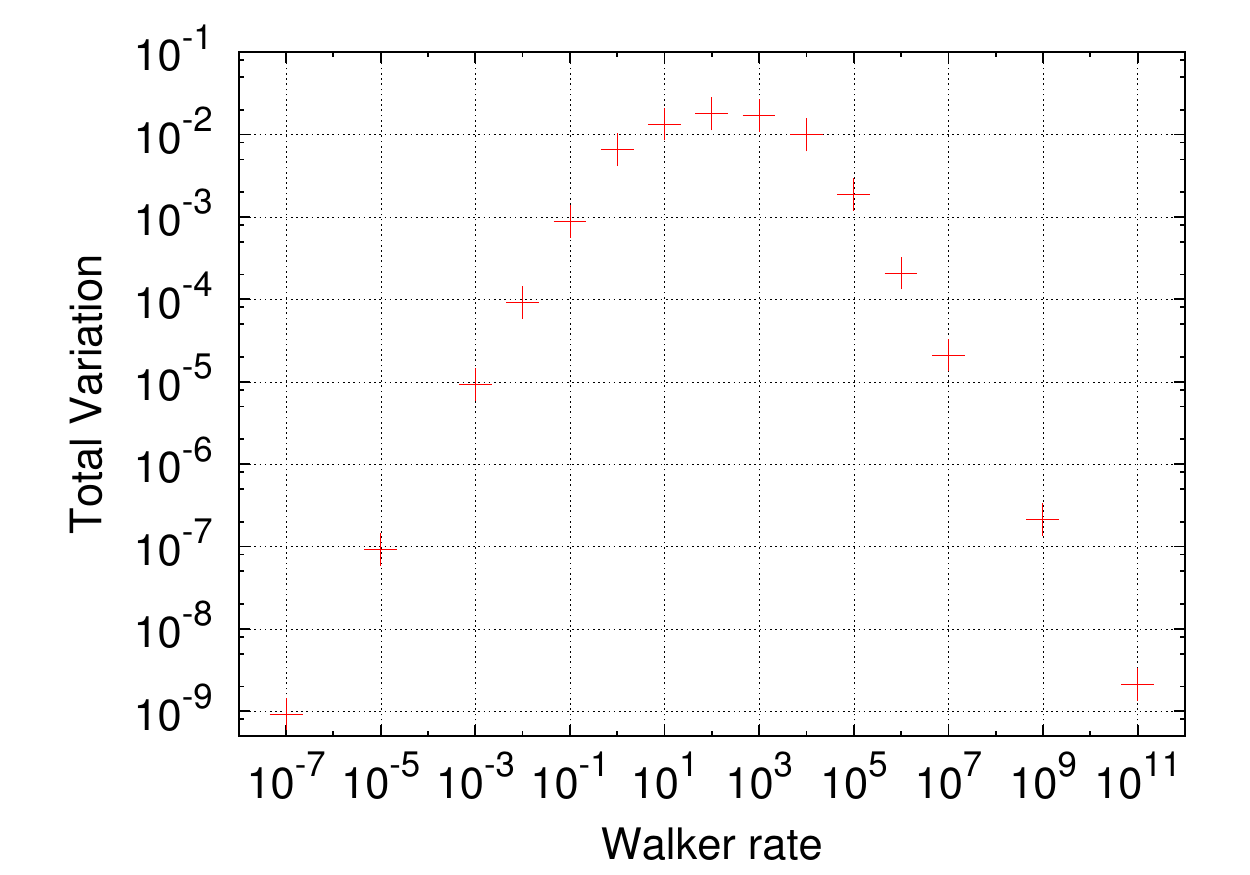}}
\caption{\label{fig:K3} Characteristics of random walks on a $K_3$ edge Markovian graph model.}
\end{figure*}

Figure \ref{fig:K3-a} shows the exact steady state distribution of the 
random walk as a function of the walker rate. 
Interestingly, while all edges have the same time average and all configurations $A_k$, $k=1,\ldots, 2^{|E|}$,
have the same $\sigma_k$, the walker steady state distribution still depends on the 
walker rate. Moreover, the behaviors of the fast and slow walkers differ. While 
the slow walker converges to a uniform distribution over the nodes, the fast walker always favors node 3, the node not incident to the fastest changing edge $(1,2)$. Despite the relatively small differences 
in the walker distribution ($P[W=3]$ varies by 7\%), the point of this example is to illustrate that such 
differences can arise even in small and simple models.

Figure \ref{fig:K3-b} also shows the maximum absolute difference between our theoretical results 
for the fast and slow walker and the actual distribution obtained exactly. In particular, we present 
the {\em total variation distance} between the two distributions, defined as $\max_{i=1,\ldots,n} | \pi(i) - \pi'(i) |$, 
where $\pi$ is the exact walker distribution and $\pi'$ is either the fast or slow walker distribution, and $n$ is the 
number of nodes in the graph. For walker rates greater than one, the theoretical results for the fast walker were 
used ($\pi'$ as defined in Section \ref{sec:fast}), while for rates smaller 
than one the results for the slow walker were used ($\pi'$ as defined in Section \ref{sec:slow_walker}). Note that as the walker slows down or speeds up, the total variation decreases (graph in log-log scale). Our numerical results indicate that the fast and slow walker become close to 
our asymptotic results as their rates increase and decrease, respectively. When the timescales of the walker and graph dynamics are similar, 
the total variation distance between the asymptotic cases and the steady state distribution is relatively larger, as expected.

Moreover, if we consider the ``degree-time distribution'' of a given node, namely, the 
fraction of time that node $v \in V$ has degree $0, 1, \ldots$, we note that all 
three nodes have identical degree-time distributions but the fraction of time the walker spends on each node varies.
This indicates that the degree-time distribution is insufficient to characterize the walker steady state.
This is also like true for transient metrics as well.

\subsubsection{Edge Markovian: 6-node kite}

\begin{figure*}[ht]
\centering
\subfloat[\label{fig:kite-a} The edge Markovian model over the 6-node kite graph where edges belong to one of two On-Off processes.]{
\includegraphics[width=2.0in,height=1.8in]{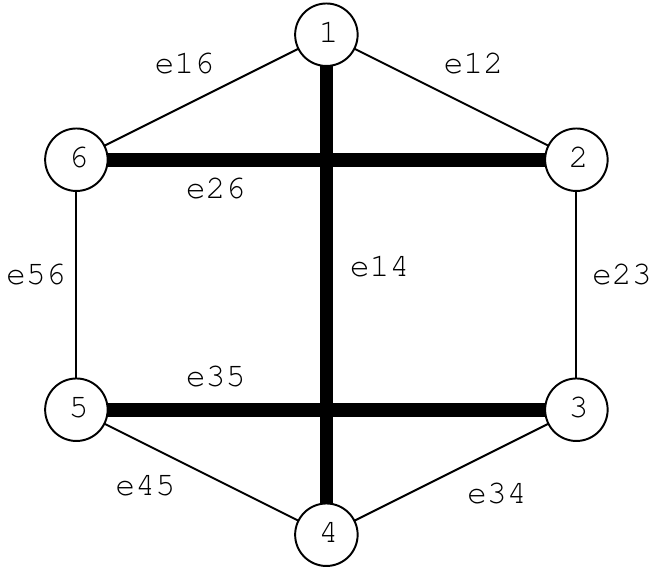}}
\hspace{1cm}
\subfloat[\label{fig:kite-b} Steady state distribution of walker as a function of walker rate (nodes 5 and 6 not shown, but are identical to nodes 2 and 3).]{
\includegraphics[width=3in]{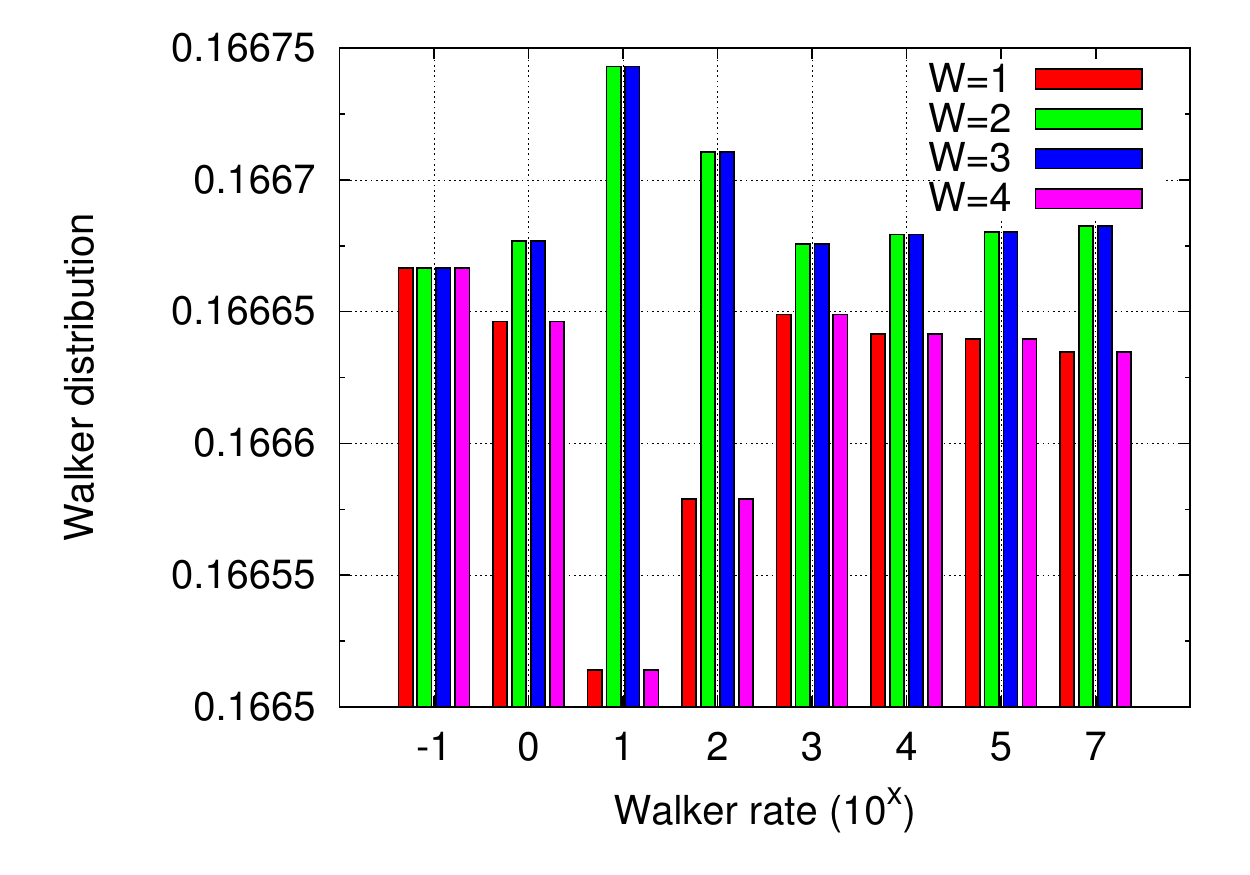}}
\caption{\label{fig:kite} Characteristics of random walks on a 6-node kite edge Markovian graph model.}
\end{figure*}

Now consider an edge Markovian process over the ``kite graph'' illustrated 
in Figure~\ref{fig:kite-a}. In particular, let all thin edges $e$ have On-Off rates $\Lambda_1(e) = \Lambda_0(e) = 1$.
Similarly, let all thick edges $e^\prime$ have On-Off rates $\Lambda_1(e^\prime) = 100, = \Lambda_0(e^\prime) = 10$. 
Note that locally all nodes are connected through identical and independent On-Off processes, two thin edges (On-Off rates equal to one) and one thick edge (On-Off rates equal to one hundred). 
Thus, in some sense, ``locally'' all nodes are indistinguishable. Surprisingly, even in this case the behavior of the walker depends on its rate, as shown in Figure \ref{fig:kite-b}.
Clearly, the structure of the graph plays an important role, as illustrated in this example, as the fast and the slow walkers have different time stationary distributions for different nodes.
 This indicates the difficulty 
of characterizing the exact behavior of random walks in general graph dynamics, even if we limit ourselves to 
the class of edge Markovian graphs.

%% file: buslines_example.tex
\subsection{A simple vehicular DTN} \label{sec:DTN}

In this section we consider an application of our modeling framework to a simple 
vehicular disruption-tolerant networks (DTN) model.
Our model captures some essential characteristics of DTNs.
Consider a set of buses equipped with 
wireless routers moving around according to their routes. Two buses establish communication 
when they are within the coverage radius of the wireless routers. Buses belonging to the 
same line (route) move from one bus stop to another following a predefined sequence of stops 
in a circular fashion.
The following notation is used to describe the model:
$S = \{s_1, \ldots, s_o\}$ is the set of bus stops across all bus 
lines, where $o$ is the number of different stops; 
$L_i \in S^{n_i}$ is a vector with the sequence of stops for bus line $i$, 
and $n_i$ is the number of stops at bus line $i$; 
$l$ denotes the number of different bus lines and $b_i$ is the number of 
different buses operating in line $i = 1, \ldots, l$;
Assume that both $\xi^i_k$, the amount of time line $i$ bus stays at stop $k$, and
$\zeta^i_{kl}$, the amount of time it takes line $i$ bus to move from stop $k$ to $l$, 
are exponentially distributed random variables, but can have different parameter values 
for any $i, k, l$. In addition, buses move independently 
of each other, including those in the same line.

The bus routes and the coverage radius of the wireless router allow two or more buses to 
exchange information when buses are at the same stop. Thus, if two or more different bus 
lines share at least one bus stop in their route, then buses from these lines 
will be able to communicate at the shared bus stops. Moreover, two or more buses 
from the same line can communicate in any stop of their line, as they can always meet at these 
stops. Finally, we assume that the communication radius is smaller than the distance between 
bus stops, such that buses only communicate when located at a shared bus stop.

\begin{figure*}[ht!]
\centering
\subfloat[\label{fig:bus-example} Example of a vehicular DTN with eleven bus stops, three bus lines and four buses (line 3 has 2 buses).]{
\includegraphics[width=2.8in,height=1.7in]{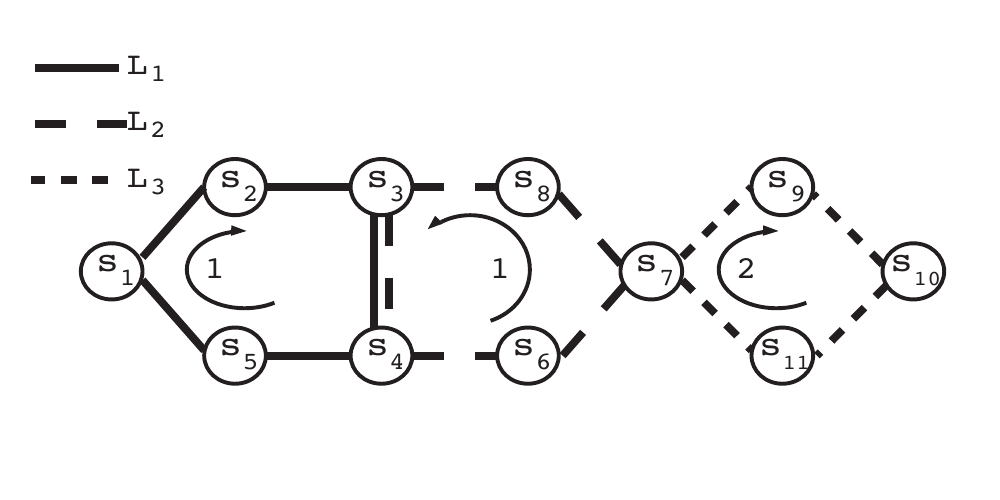}
}
\hspace{0.2cm}
\subfloat[\label{fig:bus-nets} Connectivity graphs and transitions between them according to the bus system illustrated in Figure \ref{fig:bus-example}.]{
\includegraphics[width=2.8in,height=1.7in]{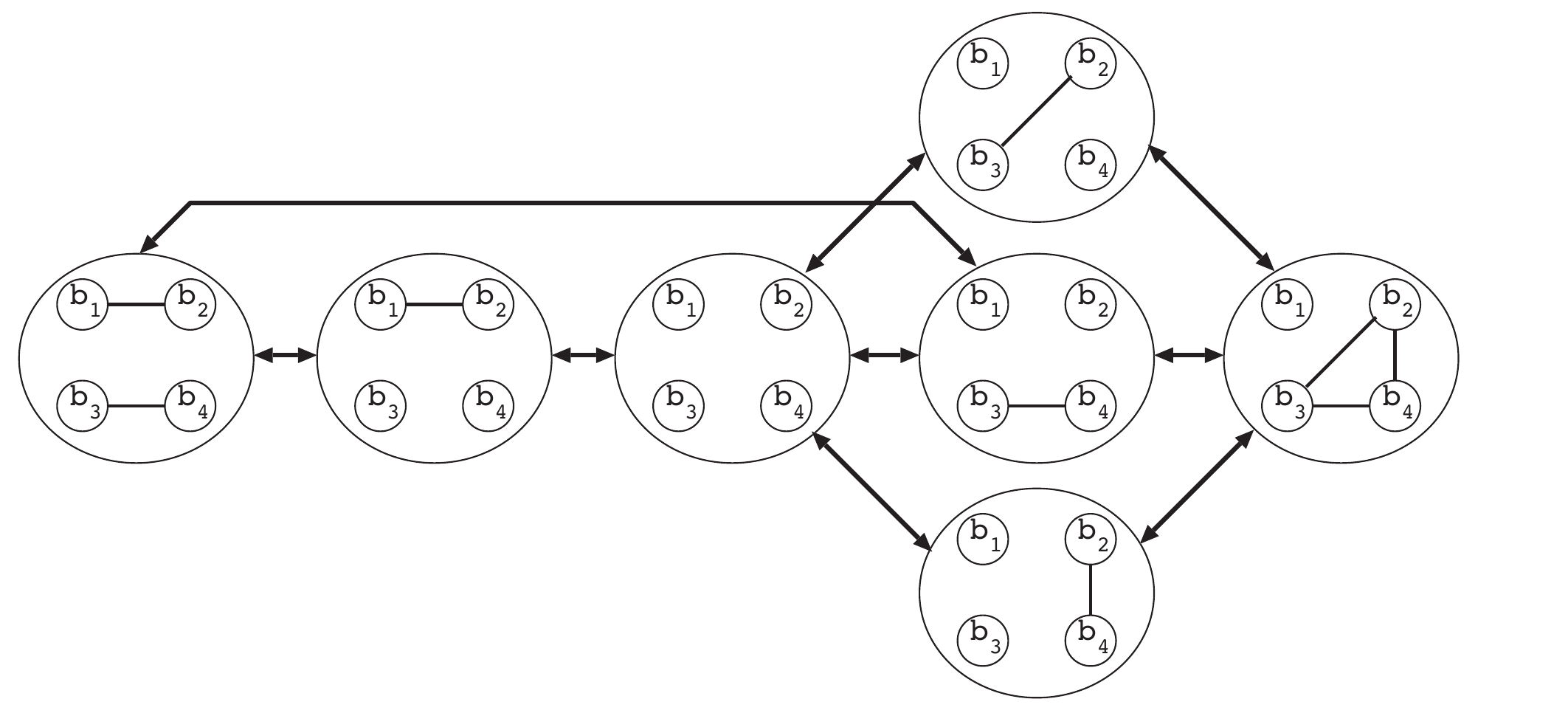}
}
\caption{\label{fig:bus} Example of a simple bus system (a) and the induced dynamic graph model (b).}
\end{figure*}

Figure~\ref{fig:bus-example} shows an example with 11 bus stops, three bus lines ($l=3$), 
defined by $L_1 = (s_1, s_2, s_3, s_4, s_5)$, $L_2 = (s_3, s_4, s_6, s_7, s_8)$ and
$L_3 = (s_7, s_9, s_{10}, s_{11})$, 
and four buses: $b_1 = 1$, $b_2 = 1$, $b_3 = 2$ (line three has two buses). Note that lines 1 and 2 
share two bus stops ($s_3$ and $s_4$) and that lines 2 and 3 share one bus stop ($s_7$). 

%

Consider a continuous-time random walker (CTRW) moving around the buses with rate $\gamma$. The 
goal is to determine the fraction of time that the walker spends in each bus or in each bus line. 
This problem can be formulated and solved using the modeling framework proposed in this paper. The first 
step is to construct a dynamic connectivity graph model from the movement of the buses in 
their respective lines. In particular, each bus is a node in the graph, since this 
corresponds to a possible location for the random walk. Moreover, each possible 
configuration of buses on their stops will define a connectivity graph, where nodes 
(buses) in the same stop are all within communication radius of one another. Note that each 
connectivity graph is composed of connected components that are all cliques (fully connected 
subgraph), since all buses in the same stop can communicate. 

Consider the example in Figure \ref{fig:bus-example} and the possible connectivity 
graphs that can be created, which are illustrated in Figure \ref{fig:bus-nets}. The connectivity 
graph has four nodes, corresponding to the four buses.
Each bus can be in a different 
stop, thus yielding a connectivity graph with no edges. Also, bus in line 2 can be at stop $s_7$ at the 
same time as the two buses from line 3, yielding a connectivity graph where these three buses are 
all connected.
Finally, note that not all graphs with four nodes are possible, since 
different lines may not have stops in common such as lines 1 and 3, for instance. 

The transitions between graph configurations are shown in 
Figure~\ref{fig:bus-nets}. 
In our model, buses cannot simultaneously leave a stop.
As a consequence, the number of allowed transitions is reduced. 
Once the dynamic graph model is constructed, 
we can obtain the state holding times for each graph configuration. 
This is possible if the holding times at bus stations and the amount of time it takes a bus
to move from a station to another are exponentially distributed. 
However, this is non-trivial in the general case, since buses can move along their 
routes without changing the connectivity graph. Moreover, totally different bus configurations over 
the set of stops can lead to the same connectivity graph. Since our modeling framework makes 
no assumption on the state (static graph) holding times of the dynamic graph model we
can extend the exponential assumption considering general distributions for the holding
times {\em at each graph configuration}, assuming only
that the expected holding time is finite for all static graphs and that dynamic graph process 
is stationary, ergodic and T-connected. 

The model constructed from this scenario matches the case studied in 
Section~\ref{sec:time_scale_inv} in which the stationary distribution of the random walk is 
time-scale invariant and uniform over the set of nodes in the graph, independent of graph 
dynamics (see Theorem \ref{thm:samefixed}). This occurs since every connected component of 
every possible graph configuration is a clique, thus, having identical degree within each 
component. Therefore, the fraction of time the walker spends in any given bus is simply 
$1/\sum_{j}b_j$, while the fraction of time spent in bus line $i$ is simply $b_i/\sum_{j}b_j$. 
Thus, CTRWs could find applications in searching for information or sampling properties 
in such systems, as its steady state does not depend on the graph dynamics.

%% file: related.tex
\section{Related work}
\label{sec:related}

Random walks have been widely used to understand and characterize graphs due 
to their well understood steady state behavior. By leveraging the steady state 
distribution of random walks, principled mechanisms for characterizing and 
estimating vertex-related properties have been devised \cite{NewmanCentrality,Gjoka10,Massoulie06,Rasti09,RT10}.

It follows that random walks can potentially be used to understand and characterize 
dynamic graphs. In fact, efforts in this direction concerning time-independent 
dynamic graphs (i.e., each snapshot is independent of the previous) have appeared 
in the literature \cite{Clementi07,Hedetniemi88,Kempe02,Pittel87}, mainly in 
the context of determining upper and lower bounds for the cover time of random 
walks. More recently, proposals to define time-dependent dynamic graph models as 
well as characterize random walks in them have also appeared in the literature
\cite{Avin08,Clementi08,Acer10}. However, these efforts have focused on 
discrete-time dynamic graph models with a goal of computing the cover time of 
random walks either in special graph structures \cite{Avin08}, in specific dynamic 
graph models \cite{Clementi08}, or through numerical evaluations \cite{Acer10}. 
Our work differs from these in the sense that we consider continuous-time 
dynamic graph models and continuous-time random walks with the goal 
of analytically characterizing the steady state behavior of the walker. 
Moreover, our prior work on this topic considered only Markovian dynamics 
and characterized the steady state behavior of the walker only under 
time-scale separation \cite{Ribeiro2011}. 

Finally, random walks have also been used as a sampling mechanism to estimate 
characteristics of vertices (e.g., fraction of vertices of a particular 
kind) in large static graphs \cite{Gjoka10,Massoulie06,RT10}. More recently, 
efforts to measure characteristics of vertices in dynamic graphs have also 
appeared in the literature \cite{Stutzbach09,Rasti09}. However, these are 
mostly preliminary and exploratory papers, indicating potential pitfalls and 
biases introduced by fast changing dynamic graphs. In contrast, our works 
is a first step at providing a theoretical foundation that can then be applied 
to estimate characteristics of vertices in dynamic graphs.

%% file: conclusion.tex
\section{Conclusion}
\label{sec:conclusion}

Understanding the long-term behavior of CTRW over dynamic graphs is an important step towards a comprehensive 
study of dynamic graphs.
Since the steady state distribution of random walks on static graphs is arguably the most important characteristic of these processes, 
it is of vital importance to characterize the CTRW steady state distribution for a broad class of dynamic graph processes.  
Unlike random walks on static graphs, CTRWs on dynamic graphs have a non-trivial behavior.
The walker rate as well as the process that governs the graph dynamics both impact the asymptotic time stationary distribution of the walker (the amount of time a walker spends on each node).

Our main results assume the graph process to be asymptotically stationary, ergodic and T-connected.
We make no independence assumptions concerning the times 
the process resides in each of the graph configurations or the particular
distribution of these residence times.  
In other words, the time spent in a given graph configuration can be dependent
on the residence time at other configurations and, in addition, the distribution of the
time the network spends in a configuration can be general.
Moreover, the dynamic graph can have temporarily disconnected nodes as long as
the dynamic graph is T-connected.

Under this general scenario, we have obtained the steady state distribution of cases in which the walker is either
much faster or much slower as compared to the rate of changes in graph configurations.
We have obtained a sufficient condition for the CTRW stationary distribution to be
invariant to the walker rate and presented examples that illustrate models in which
these results are applicable.
In this context, additional application examples for the constant degree constraint
can be found in P2P networks where peers maintain a nearly fixed number of connections.
Hence, Theorem~\ref{thm:samefixed} helps explain why sampling these dynamic 
networks using random walks leads to meaningful results~\cite{Stutzbach09}. 

The examples in this paper serve mainly for illustrative purposes.
However, we believe that the theory we developed will find applications in many important areas.
Examples of promising application areas are in sampling DTN and P2P networks.

%% file: appendix.tex

\appendix

\begin{subappendices}

\section{Proof of Proposition~\ref{prop:AtSE}}\label{stationarity}

From a classical construction of ergodic theory (see e.g. \cite[Prop.\ 11.4]{Robert}), we know that
there exist a probability space $(\Omega, {\cal F}, P)$ and
ergodic endomorphisms  $(\theta^t)_{t\in \mathbb{R}}$, $\theta^t : \Omega\to \Omega$ - called a  flow - on this probability space
on which the stationary sequence
$\Psi=\{X_n,S_n\}_{n\in \mathbb{Z}}$ is defined, such that 
\begin{equation}
\label{stat-sequence}
S_n=S \comp \hat\theta^n,\quad X_n=X\comp \hat\theta^n,\quad n\in \mathbb{Z},
\end{equation}
with $(S,X)$ rvs distributed as $(S_n,X_n)$ and $\hat \theta := \theta^{S_0}$.

Observe that  the stationary sequence defined in (\ref{stat-sequence}) satisfies 
\begin{equation}
\label{ergo-sequence}
\lim_{N\to\infty}\frac{1}{N}\sum_{n=1}^N f(X_n,S_n) = \lim_{N\to\infty}\frac{1}{N}\sum_{n=1}^N f(X,S)\comp \hat\theta^n
= E[f(X,S)]\quad \hbox{a.s.}
\end{equation}
for any nonnegative measurable mapping $f$ on $\Omega$, that is, it is an ergodic sequence.
The second equality in (\ref{ergo-sequence}) follows from the ergodicity of the flow  $(\theta^t)_{t\in \mathbb{R}}$.

Furthermore, for any $n\in \mathbb{Z}$,  $\omega\in \Omega$,
\begin{equation}
\label{shift}
T_n(\theta^s(\omega))=T_{n+k}(\omega)-s,
\end{equation}
where $k$ is the unique integer such that $T_k(\omega)\leq s<T_{k+1}(\omega).$\footnote{In words, (\ref{shift}) says that for $\omega=(s_n,s_n)_{n\in \mathbb{Z}}\in \Omega$,
the $n$th point of the point process $(t_n)_{n\in \mathbb{Z}}$ (with $t_{n+1}-t_n=s_n$)
to the right (resp. left) of $t=0$ when the trajectory is shifted by $s$ (i.e. $x_n$ is replaced by $x_n-s$ for
every $n\in  \mathbb{Z}$) is the $(n+k)$-th point to the right (resp. left) of $t=0$ if $t_k\leq s<t_k$.}
A more compact notation for  (\ref{shift}) is $T_n \comp \theta^s=  T_{n+k}-s$ if $T_k\leq s<T_{k+1}$.

We are now in position to prove the stationarity of $\{A(t)\}_{t\in \mathbb{R}}$.

Recall that $A(t)=\sum_{n\in Z} X_n {\bf 1}(T_n\leq t <T_{n+1})$. We have
\[
A(t) \comp \theta^s = \sum_{n\in Z} X_n {\bf 1}(T_n \comp \theta^s\leq t <T_{n+1} \comp \theta^s).
\]

If $T_k\leq s<T_{k+1}$, by (\ref{shift}),
\begin{eqnarray*}
A(t) \comp \theta^s &=& \sum_{n\in Z} X_n {\bf 1}(T_{n+k} \leq t +s <T_{n+1+k})\\
&=& \sum_{n\in Z} X_{n-k} {\bf 1}(T_{n} \leq t +s <T_{n+1})\\
&=& \theta^{T_k}  \sum_{n\in Z} X_{n} {\bf 1}(T_{n} \leq t +s <T_{n+1})\\
&=& \theta^{T_k} \comp A(t+s),
\end{eqnarray*}
where we have used the relation $X_{n-k}=X_n\comp \theta^{T_k}$.
The above can be rewritten as
\[
A(t)\comp \theta^s =  A(t+s)\comp  \sum_{k\in Z} \theta^{T_k}  {\bf 1}(T_k\leq s <T_{k+1})
\]
or, equivalently,
\[
A(t+s) = A(t) \comp \sum_{k\in Z} \theta^{s-T_k}  {\bf 1}(T_k\leq s <T_{k+1}).
\]
This shows that $\{A(t)\}_t$ is stationary since
\begin{eqnarray*}
P(A(t+s)\in C)&=& P\left(\sum_{k\in Z}A(t)\comp  \theta^{s-T_k}  {\bf 1}(T_k\leq s <T_{k+1}) \in C\right)\\
&=&\sum_{k\in Z} P( A(t)\comp \theta^{s-T_k} \in C, {\bf 1}(T_k\leq s <T_{k+1}))\\
&=&\sum_{k\in Z} P( A(t)\in C,  {\bf 1}(T_k\leq s <T_{k+1}))\\
&=&P(A(t) \in C).
\end{eqnarray*}
In particular,
\[
A(s)=A(0)\comp \sum_{k\in Z} \theta^{s-T_k}  {\bf 1}(T_k\leq s <T_{k+1}).
\]
\hfill\done

%
%
%
%
%
%

\section{Proof of Proposition~\ref{prop:uniform}.}\label{uniform}

Fix $\epsilon>0$. All rvs are defined on the probability space $(\Omega, {\cal F},P)$.
By Egorov's theorem\footnote{Egorov's theorem applies here since $\Omega$ has a finite measure w.r.t.\
$P$ as $P(\Omega)=1$.}~\cite[pg.\ 43, Theorem 2]{Kolmogorov} we know that there exists 
a Borel set ${\cal C}_\epsilon\subset {\cal F}$ with $P({\cal C}_\epsilon)<\epsilon$,
such that the convergence
in Theorem~\ref{teo:birkhoff} is uniform on $\Omega-{\cal C}_\epsilon$, namely, there exists $T_\epsilon$
such that  (we write $A(x,\omega)$ for $A(x)$ to emphasize that $A(x)$ 
is a P-measurable rv)
\begin{equation}
\label{app-int1}
\sigma_k-\epsilon < \frac{1}{t}\int_0^t {\bf 1}_{\{A(x,\omega)=A_k\}}dx   <\sigma_k+\epsilon
\end{equation}
for all $t>T_\epsilon$ and for all $\omega \in \Omega-{\cal C}_\epsilon$.

Let ${\cal B}\in {\cal F}$ with $P({\cal B})>0$. Integrating w.r.t.\ $dP(\omega)$ for $w\in {\cal B}-{\cal C}_\epsilon$ 
in (\ref{app-int1}) gives
\begin{equation}
(\sigma_k-\epsilon) P({\cal B}-{\cal C}_\epsilon) < \int_{{\cal B}-{\cal C}_\epsilon} dP(\omega) 
\frac{1}{t}\int_0^t {\bf 1}_{\{A(x, \omega))=A_k\}} dx  <(\sigma_k+\epsilon)P({\cal B}-{\cal C}_\epsilon)
\label{app-int2}
\end{equation}
for all $t>T_\epsilon$ and for all $\omega \in {\cal B}-{\cal C}_\epsilon$.

By Fubini's theorem
\begin{eqnarray}
\int_{{\cal B}-{\cal C}_\epsilon} \int_0^t {\bf 1}_{\{A(x, \omega)=A_k\}} dx dP(\omega)  &=&
\int_0^t \int_\Omega {\bf 1}_{\omega \in {\cal B}-{\cal C}_\epsilon}  {\bf 1}_{\{A(x, \omega)=A_k\}}
dP(\omega)dx\nonumber\\
&=&\int_0^t  P(A(x)=A_k\cap ({\cal B}-{\cal C}_\epsilon))dx.
\label{app-int3}
\end{eqnarray}
Since 
\begin{eqnarray}
P(A(x)=A_k \cap ({\cal B}-{\cal C}_\epsilon))&=&
 P(A(x)=A_k \cap {\cal B})- P(A(x)=A_k \cap {\cal C}_\epsilon)\nonumber\\
&=& P(A(x)=A_k| {\cal B})P({\cal B})- P(A(x)=A_k | {\cal C}_\epsilon)P({\cal C}_\epsilon),
\label{app-int4}
\end{eqnarray}
we obtain from (\ref{app-int2})-(\ref{app-int4}) that 
\begin{eqnarray}
(\sigma_k-\epsilon) P({\cal B}-{\cal C}_\epsilon) + P({\cal C}_\epsilon)f(t,\epsilon)&<&\frac{P({\cal B})}{t}\int_0^t
P(A(x))=A_k| {\cal B}) dx\nonumber\\
&<& (\sigma_k+\epsilon) P({\cal B}-{\cal C}_\epsilon) + P({\cal C}_\epsilon)f(t,\epsilon)
\label{app-int5}
\end{eqnarray}
for all $t>T_\epsilon$, with $f(t,\epsilon):= (1/t)\int_0^t P(A(x)=A_k| {\cal C}_\epsilon) dx$.

Dividing (\ref{app-int5}) by $P({\cal B})$ (recall that $P({\cal B})>0$) gives
\begin{eqnarray}
(\sigma_k-\epsilon) \frac{P({\cal B}-{\cal C}_\epsilon)}{P({\cal B})} + \frac{P({\cal C}_\epsilon)}{P({\cal B})}f(t,\epsilon)
&<&\frac{1}{t}\int_0^t
P(A(x)=A_k| {\cal B}) dx\nonumber\\
&<& (\sigma_k+\epsilon)  \frac{P({\cal B}-{\cal C}_\epsilon)}{P({\cal B})} + \frac{P({\cal C}_\epsilon)}{P({\cal B})}f(t,\epsilon).
\label{app-int6}
\end{eqnarray}
As $\epsilon\to 0$, (i) $\frac{P({\cal C}_\epsilon)}{P({\cal B})}f(t,\epsilon)\to 0 $ since
$0\leq \frac{P({\cal C}_\epsilon)}{P({\cal B})}f(t,\epsilon)<\frac{\epsilon}{P({\cal B})}$ from the definition of ${\cal C}_\epsilon$ and the fact that $0\leq f(t,\epsilon)\leq 1$.
On the other hand, (ii) both terms $(\sigma_k\pm\epsilon)  \frac{P({\cal B}-{\cal C}_\epsilon)}{P({\cal B})} $ go to $\sigma_k$  
as $\epsilon\to 0$ since $\lim_{\epsilon \to 0} P({\cal B}-{\cal C}_\epsilon)/P({\cal B})=
\lim_{\epsilon \to 0}(P({\cal B})-P({\cal C}_\epsilon))/P({\cal B})=1$. 
Since $T_\epsilon \to \infty$ as $\epsilon \to 0$ we conclude from (i)-(ii) above and from (\ref{app-int6})
that $\lim_{t\to\infty} (1/t)\int_0^t P(A(x)=A_k)|{\cal B}) dx=\sigma_k$ for all Borel sets ${\cal B}$ 
such that $P({\cal B})>0$.\hfill\done

\section{Proof: {\large $T$} is finite} \label{largeT}
The graph process $\{A(t)\}_{t \in \mathbb{R}}$ is time stationary and ergodic. It follows from Proposition~\ref{prop:uniform}  that  there exists $T_k > 0$ s.t.
\[
   \frac{1}{t} \int_{x}^{x+t} {\bf 1}(A(s) = A_k) ds  > \sigma_k/2 \: , \quad x \geq 0,\; t\ge T_k
\]
independent of $x$.  Choose $k_0 =  {\operatorname{argmin}}_k \{\sigma_k\}$ and $T' =\max \{T_k\}$.  Thus
\[
    \frac{1}{T'} \int_{x}^{x+T'} {\bf 1}(A(s) = A_k) ds  > \sigma_{k_0}/2 \: , \quad x \geq 0 .
\]
Consequently the graph process spends at least $\sigma_{k_0} T'/2$ units of time in configuration $k$ during interval $[x,x+T')$ regardless of the state of the graph process at $t=x$. Now consider the situation where $U_1(x) = u$, $U_2(x) = w$, and $A(x) = A_k$.  Let $A_k = A_{l_1}, A_{l_2}, \ldots , A_{l_j}$ be a sequence of graphs such that there is a temporal path between $u$ in $A_{l_1}$ and $w$ in $A_{l_j}$.  This requires that there be paths within each graph configuration, $p_i = (v_{i,0}, v_{i,1}, \ldots ,v_{i,n_i})$ that satisfy $v_{1,0} = u$, $v_{i+1,0} = v_{i,n_{i}}$, $i=1,\ldots , j-1$, and $v_{j,n_j} = w$.  Let $h_{u,w}$ denote the number of physical hops on this path and $j = H_{u,w}$ the number of configurations in the sequence. We focus now on the event that walker 1 progresses from node $u$ to node $w$ in the interval $[x, x+jT')$ $jT'$ by progressing across path $p_i$ during $[x+(i-1)T', x+iT')$ while walker 2 remains at node $w$. The probability of this event, $p_{u,w}$ is bounded from below by 
\baqm
 \lefteqn{p_{u,w}  \ge   e^{-\gamma_{\max} H_{\max}T'} e^{-\gamma_{\max} H_{\max}T'(1-\sigma_{k_0}/2)}} \\
& & \mbox{} \times
\bigl(\frac{1}{d_{\max}}\bigr)^{h_{\max}}\prod_{i=1}^j P\bigl(\sum_{\ell =0}^{n_i-1} Z_{i,v_\ell} < T'\sigma_{k_0}/2 \le \sum_{\ell =0}^{n_i} Z_{i,v_\ell}\bigr)
\eaqm
Here $H_{\max} = \max_{u,w} H_{u,w}$, $h_{\max} = \max_{u,w} h_{u,w}$, and $Z_{i,v}$ denotes the time between a walker arriving to node $v$ in configuration $A_{l_i}$ and taking its next step.  This time is exponentially distributed with rate $\gamma_{l_i,v}$ and there exists some $q>0$ such that 
\[
P\bigl(\sum_{\ell =0}^{n_i-1} Z_{i,v_\ell} < T'\sigma_{k_0}/2 \le \sum_{\ell =0}^{n_i} Z_{i,v_\ell}\bigr) > q 
\]
for all $u,w$.  Hence 
\[
p_{u,w} \ge p_0 \equiv e^{-2\gamma_{\max} H_{\max}T'} 
\bigl(\frac{1}{d_{\max}}\bigr)^{h_{\max}} q
\]
Last, $T_0 \equiv H_{\max} T'$.

\section{Irreducibility of $\bP$} \label{Pirred}
Consider $\bP$ as defined in~\eqref{eq:P} to be the adjacency matrix of a weighted directed graph $G$ with self-edges.
Then by~\cite[Theorem 6.2.24]{Horn} $\bP$ is irreducible if $G$ is strongly connected (as the elements of $\bP$ are finite, i.e., $\Vert \bP \Vert_\infty < \infty$). 
What we need to show is that $G$ is strongly connected.
Let 
$$
\epsilon = \frac{\gamma}{\gamma^\prime}  \min_{k=1,\ldots,m, \, \forall u \in V}\left(\frac{\sigma_k}{d_{k,u}}\right) \, ,
$$
where $\gamma$ and $\gamma^\prime$ as defined in~\eqref{eq:P}.
Clearly $\epsilon > 0$.
Now decompose $\bP$ into two parts: $\bP = \epsilon \sum_{k=1}^m A_k + \Xi$.
From the definition of $\bP$,~\eqref{eq:P}, it is clear that $\Xi \geq 0$.
By the definition of T-connectivity (Definition~\ref{def:Tc}) the (undirected weighted) graph with adjacency matrix $A = \sum_{k=1}^m A_k$ is connected.
Thus, the graph with adjacency matrix $A_\epsilon = \epsilon \sum_{k=1}^m A_k$ (remember that $\epsilon > 0$) must also be connected.
As the adjacency matrix of $G$ can be written as $A_\epsilon + \Xi$, $\Xi \geq 0$, then $G$ is strongly connected, finishing our proof.

%
%
%
%
%

\end{subappendices}

%% file: paper.bbl
\begin{thebibliography}{10}

\bibitem{Acer10}
Utku~G\"{u}nay Acer, Petros Drineas, and Alhussein~A. Abouzeid.
\newblock Random walks in time-graphs.
\newblock In {\em MobiOpp}, pages 93--100, 2010.

\bibitem{Avin08}
C.~Avin, M.~Koucky, and Z.~Lotker.
\newblock How to explore a fast-changing world. (on the cover time of dynamic
  graphs).
\newblock In {\em ICALP}, pages 121--132, 2008.

\bibitem{Clementi08}
A.~Clementi, C.~Macci, A.~Monti, F.~Pasquale, and R.~Silvestri.
\newblock Flooding time in edge-markovian dynamic graphs.
\newblock In {\em PODC}, pages 213--222. ACM, 2008.

\bibitem{Clementi07}
A.~Clementi, F.~Pasquale, A.~Monti, and R.~Silvestri.
\newblock Communication in dynamic radio networks.
\newblock In {\em PODC}, pages 205--214. ACM, 2007.

\bibitem{Franken}
P.~Franken, D.~Konig, U.~Arndt, and V.~Schmidt.
\newblock {\em Queues and Point Processes}.
\newblock J. Wileyl, 1982.

\bibitem{Gjoka10}
M.~Gjoka, M.~Kurant, C.~Butts, and A.~Markopoulou.
\newblock A walk in {Facebook}: Uniform sampling of users in online social
  networks.
\newblock In {\em INFOCOM}, 2010.

\bibitem{Hedetniemi88}
S.M. Hedetniemi, S.T. Hedetniemi, and A.L.Liestman.
\newblock A survey of gossiping and broadcasting in communication networks.
\newblock {\em Networks}, 18(4):319--349, 1988.

\bibitem{Horn}
R.~Horn and C.~Johnson.
\newblock Matrix analysis.
\newblock Technical report, Cambridge University Press, 1985.

\bibitem{Kempe02}
D.~Kempe and J.~Kleinberg.
\newblock Protocols and impossibility results for gossip-based communication
  mechanisms.
\newblock In {\em FOCS}, pages 471--480, 2002.

\bibitem{Kolmogorov}
A.~N. Kolmogorov and S.~V. Fomin.
\newblock {\em Elements of the Theory of Functions and Functional Analysis},
  volume~2.
\newblock Dover, 1957.

\bibitem{Massoulie06}
Laurent Massouli{\'e}, Erwan {Le Merrer}, Anne-Marie Kermarrec, and Ayalvadi
  Ganesh.
\newblock Peer counting and sampling in overlay networks: random walk methods.
\newblock In {\em PODC}, pages 123--132, 2006.

\bibitem{moyal}
P.~Moyal.
\newblock A generalized backwards scheme for solving non monotonic stochastic
  recursions.
\newblock {\em Arxiv preprint arXiv:1009.1235}, Sept. 2010.

\bibitem{NewmanCentrality}
M.E.J. Newman.
\newblock A measure of betweenness centrality based on random walks.
\newblock {\em Social Networks}, 27(1):39--54, 2005.

\bibitem{Pittel87}
B.~Pittel.
\newblock On spreading a rumor.
\newblock {\em SIAM Journal on Applied Mathematics}, 47(1):213--223, 1987.

\bibitem{Rasti09}
A.~H. Rasti, M.~Torkjazi, R.~Rejaie, N.~Duffield, W.~Willinger, and
  D.~Stutzbach.
\newblock Respondent-driven sampling for characterizing unstructured overlays.
\newblock In {\em INFOCOM Mini-Conference}, 2009.

\bibitem{RT10}
B.~Ribeiro and D.~Towsley.
\newblock Estimating and sampling graphs with multidimensional random walks.
\newblock In {\em SIGCOMM IMC}, Nov 2010.

\bibitem{Ribeiro2011}
Bruno Ribeiro, Daniel Figueiredo, Edmundo de~Souza~e Silva, and Don Towsley.
\newblock Characterizing continuous-time random walks on dynamic networks.
\newblock {\em SIGMETRICS}, 39(1):343--344, June 2011.

\bibitem{Robert}
P.~Robert.
\newblock {\em Stochastic Networks and Queues}, volume~52 of {\em Applications
  of Mathematics}.
\newblock Springer-Verlag, 2003.

\bibitem{Shiryaev}
Albert~N. Shiryaev.
\newblock {\em Probability}.
\newblock Springer, 2nd edition, 1995.

\bibitem{Stutzbach09}
Daniel Stutzbach, Reza Rejaie, Nick Duffield, Subhabrata Sen, and Walter
  Willinger.
\newblock On unbiased sampling for unstructured peer-to-peer networks.
\newblock {\em IEEE/ACM Trans. Netw.}, 17:377--390, April 2009.

\end{thebibliography}
